\documentclass[11pt,a4paper]{article}
\pdfoutput=1
\usepackage{jcappub}
\usepackage{subfig}
\usepackage{caption}
\usepackage{hyperref}
\usepackage{amsmath}
\usepackage{mathtools}
\usepackage{tablefootnote}
\usepackage{soul}
\usepackage{ulem}
\usepackage{color}
\usepackage[usenames,dvipsnames,svgnames,table]{xcolor}
\usepackage{bm}
\usepackage{multirow}
\usepackage{graphicx}

\definecolor{darkred}{RGB}{135,0,0}

\title{Beware of commonly used approximations II: estimating systematic biases in the best-fit parameters}

\author[a,b,c]{Jos\'e Luis Bernal,}
\emailAdd{jbernal2@jhu.edu}

\author[b,c]{Nicola Bellomo,}
\emailAdd{nicola.bellomo@icc.ub.edu}

\author[d,b]{Alvise Raccanelli,}
\emailAdd{alvise.raccanelli@cern.ch}

\author[b,e]{Licia Verde}
\emailAdd{liciaverde@icc.ub.edu}

\affiliation[a]{Department of Physics and Astronomy, Johns Hopkins University, 3400 North Charles Street, Baltimore, Maryland 21218, USA}
\affiliation[b]{ICC, University of Barcelona, IEEC-UB, Mart\'i  i Franqu\`es, 1, E-08028 Barcelona, Spain.}
\affiliation[c]{Dept. de  F\'isica Qu\`antica i Astrof\'isica, Universitat de Barcelona, Mart\'i  i Franqu\`es 1, E-08028 Barcelona, Spain.}  
\affiliation[d]{Theoretical Physics Department, CERN, 1 Esplanade des Particules, CH-1211 Geneva 23, Switzerland.}
\affiliation[e]{ICREA, Pg. Lluis Companys 23, Barcelona, E-08010, Spain.} 

\abstract{
Cosmological parameter estimation from forthcoming experiments promise to reach much greater precision than current constraints.
As statistical errors shrink, the required control over systematic errors increases. Therefore, models or approximations that were sufficiently accurate so far, may introduce significant systematic biases in the parameter best-fit values and jeopardize  the robustness of cosmological analyses. We generalize previously proposed expressions to estimate \textit{a priori} the systematic error introduced in  parameter inference due to the use of insufficiently good approximations in the computation of the observable of interest or the assumption of an incorrect underlying model. Although this methodology can be applied to measurements of any scientific field, we illustrate its power by studying the effect of  modeling the angular galaxy power spectrum incorrectly. We also introduce \texttt{Multi\_CLASS}, a new, public modification of the Boltzmann code \texttt{CLASS}, which includes the possibility to compute angular cross-power spectra for two different tracers.
We find that significant biases in most of the cosmological parameters are introduced if one assumes the Limber approximation or neglects lensing magnification in modern galaxy survey analyses, and the effect is in general larger for the multi-tracer case, especially for the parameter controlling primordial non-Gaussianity of the local type, $f_{\rm NL}$.
}

\begin{document}

\begin{flushright}
CERN-TH-2020-054
\end{flushright}

\maketitle

\section{Introduction}
\label{sec:intro}
An unprecented experimental and theoretical effort during the last decades has established the $\Lambda$-Cold Dark Matter ($\Lambda$CDM) model as the standard model of cosmology, because of its striking precision in fitting most of the  available observations. This effort has brought forward percent-level constraints on some of the parameters of the model (see e.g.,~\cite{Planck18_pars,alam_bossdr12,Scolnic_pantheon,DES_cluslens}). Nonetheless, there are still some remaining tensions between experiments (see e.g.,~\cite{Verde_KITP,RiessH0_19,Wong_holicow19,Kids_Vikings}) that, in the absence of non-accounted for systematic errors, might hint at  the need for an extension of $\Lambda$CDM. 
 Forthcoming and future galaxy surveys are expected to push the envelope of observational cosmology on the large scale structure side~\cite{EMU,desi,euclid,abell:lsstsciencebook,redbook,spherex}, but significant improvements are also expected for CMB experiments~\cite{CLASS_EEforecast,SimonsObservatory,CMBS4}. Moreover, the advent of line-intensity mapping experiments~\cite{Kovetz_IMstatus,Kovetz_IM2020} with its potential for cosmology (see e.g.,~\cite{Bernal_IM}) promises to open a window to explore the Universe at higher redshift and close the gap  between CMB and galaxy surveys observations (see e.g.,~\cite{Bernal_IM_letter,Munoz_vao}).

The aim of these experiments is to achieve subpercent-level precision for parameter inference within $\Lambda$CDM and to further constrain (and possibly even detect) deviations from $\Lambda$CDM. The promising capabilities of these experiments will make possible a dramatic reduction of the statistical errors: expected statistical uncertainties are well below the current systematic error budget. Therefore, it is of crucial importance to maintain systematic biases below the statistical errors as well as a correct assessment of the final, true uncertainties.  Accurate  modeling of the target observables is one of the key ingredients needed in order to succeed in this challenge.  
 However, higher accuracy often requires more complicated and expensive modeling. This can be mitigated by introducing careful approximations in the modeling of the signal and their covariances to speed up calculations without significant reductions in accuracy. 

Using insufficiently accurate approximations in the analysis affects the inferred model parameters in two ways. First, it may shift the point in  parameter space where the posterior distribution peaks, which would introduce a systematic bias in the  best-fit parameters. Second, the shape of the posterior distribution may be affected, yielding a misestimate of the uncertainties. In summary, these two effects can be understood as errors in the parameters values and  in their error-bars. Therefore, the effects on the posterior (in terms of potential shifts of the best-fit parameters and their errors) of any approximation to be adopted must be estimated quantitatively in light of the forecasted  experimental performance. 
 Moreover, a theoretical systematic error might be introduced even if the modeling of the observable is accurate enough, when the underlying cosmological model assumed is incorrect: the inferred cosmological parameters assuming $\Lambda$CDM will be most likely biased if the Universe is  better described by a different model.

We approach the estimation of the bias in the best-fit parameters  in the same spirit of previous works focusing on specific problems in cosmology. Some studies focused on the bias introduced by assuming an incorrect cosmological model $M_1$ instead of the correct one $M_0$, where $M_1$ and $M_0$ are nested models~\cite{Heavens_modelbias}. The general expression for nested models can be found in Refs.~\cite{Taylor_DEshear, Duncan_magnification}, where it  was used to estimate the impact of not marginalizing over a subset of nuisance parameters.  Others investigated the bias arising from an incorrect modeling of  the target observable for several specific cases, including the impact of inhomogeneous reionization on CMB anisotropies~\cite{Knox_inhomoreio}, the effect on weak lensing measurements from incorrect modeling of the non linear power spectrum and galaxy redshift distribution~\cite{Huterer_WLbias, Huterer_WLbiasPknonlin}, observational systematics~\cite{Huterer_WLbiasobs, Amara_biasWL} or baryonic feedback~\cite{Natarajan_lensingbarfeedback}, SN magnitudes \cite{Kim:2004}, the modeling of redshift-space distortions of the galaxy power spectrum~\cite{Taruya_TNS}, emission-line galaxy power spectrum measurements without including the contamination due to line interlopers~\cite{Pullen_interloperbias}, the contribution of relativistic effects on galaxy surveys~\cite{Camera_GRfnlbias, Cardona_magnification, Lorenz_GRbias, Cizmek_magnification}, and the neutrino-induced scale-dependence of the galaxy bias \cite{Raccanelli_mnu}. There are also studies focused on how parameter inference is affected by misestimations of the covariance matrices (see e.g.,~\cite{Taylor_precisioncovmat,Kodwani_lensingcosmocovmat}) or the use of an invalid likelihood (see e.g.,~\cite{Sellentin_arbitrarycovmat, Wang_NGPkLkl}). 

Instead, in this work we aim to generalize the methodology to be applicable to any observable, not necessarily in cosmology. There are two different causes that lead to parameter biases: the assumption of an incorrect model (e.g., assuming a cosmological constant in the case dark energy is dynamical) and the use of an incorrect or incomplete modeling of the observable considered (e.g., using an inaccurate modeling of the redshift-space distortions). Although the previous studies mentioned above considered only one of them, these two causes may be interdependent. Our approach treats both of them on the same footing, accounting for any possible interaction between them. 

 Our approach can be summarized as follows. We consider data drawn from an underlying model.  Then, we expand the theoretical prediction of the observable, computed according to  a given model under study,  around a fiducial set of parameters. Note that the model under study is not necessarily the same model the data are drawn from.  Subsequently, we apply this expansion to the likelihood of the observable, maximize such likelihood, and obtain the best-fit value of the parameters. We follow this procedure in two cases: a correct description of the observable and an approximated one. From here, we estimate the systematic bias comparing the best-fit parameters obtained in both cases. 
 
 This technique, as is the case for  Fisher matrix analyses, is conceived to be used prior to obtaining data. In this case, it is required to choose a model as a good representation of reality and compute the data according to it. Employing our approach before the data is obtained will allow for the determination of the level of accuracy needed in the data analysis of an experiment during its design. Moreover, it will help to quantify the significance of new contributions or corrections to the theoretical modeling of the signal in  light of the potential of future experiments. Nonetheless, the methodology can also be applied to actual observations to ascertain whether a more numerically intensive modeling is needed.

We demonstrate the power of our approach by applying it to the angular galaxy power spectrum. We primarily focus on the systematic shifts of the best-fit parameters, whereas the misestimation of the uncertainties in  parameter inference is studied in a companion paper~\cite{Bellomo_Fisher}, hereinafter Paper I.  
 Besides  traditional, single-tracer analysis, we also consider multi-tracer analyses of the galaxy clustering.  We present two practical examples: we explore how neglecting cosmic magnification or using the Limber approximation may introduce significant systematic biases ($\gtrsim 1-2\sigma$ in most of the parameters and cases under study) in the analyses of next-generation galaxy surveys. 
 
We also present and publish \texttt{Multi\_CLASS}. \footnote{\texttt{Multi\_CLASS} will be publicly available  in \url{https://github.com/nbellomo/Multi_CLASS} upon the acceptance of this work.} \texttt{Multi\_CLASS} is a modification of the  public Boltzmann code CLASS~\cite{Lesgourgues:2011re,Blas:2011rf} that allows for the computation of the angular  cross-power spectra of two different tracers of the underlying density field, with their corresponding different redshift dependence of the tracer characteristics (bias, magnification bias, evolution bias and number density distribution). This possibility enables comprehensive theoretical multi-tracer analyses.  Furthermore, it includes an implementation for primordial non-Gaussianities of the local type, parametrized by the $f_{\rm NL}$ parameter, on the clustering observables.  More details about the options and implementation of  \texttt{Multi\_CLASS}  can be found in Appendix~B of  Paper I. 

This paper is organized as follows: we generalize the methodology to estimate the bias introduced in best-fit parameters by an inaccurate modeling of a given measurement in Section~\ref{sec:shifts}; introduce our test-case, the angular galaxy power spectrum, and our  assumptions for the demonstration of the potential of our approach in Section~\ref{sec:observable}; show the estimated systematic bias introduced in  cosmological parameter inference by using the angular galaxy power spectrum without modeling lensing magnification or applying the Limber approximation in Section~\ref{sec:results}; and conclude in Section~\ref{sec:conclusions}. Appendix~\ref{app:shifts_beyond} contains an evaluation of the performance of the estimation of the bias, the derivation of the estimated bias expanding the observable up to second order on the model parameters, and a discussion of likelihoods of Wishart-distributed variables. Appendix~\ref{app:ellmin} quantifies the dependence of the bias in the best-fit parameters on the largest scales included in the angular galaxy power spectrum analysis.

\section{Systematic shift of the best-fit parameters due to incorrect modeling}
\label{sec:shifts}
We begin by defining notation and conventions. We use vector operators for quantities in parameter space; operations involving the observable space (e.g., the data vector or its covariance) are explicitly written down as sums over the matrix elements. Unless otherwise stated, all vectors are column vectors.  For the sake of clarity, the meaning of all symbols, superscripts and subscripts used in this section can be consulted in Table~\ref{tab:symbols}. 

 Let us consider a generic model $M$ specified by a set of parameters $\bm{\theta}$  that, according to some theoretical modeling, determine the observable $\Psi$ and its covariance ${\rm Cov}$. While in principle there is no need to distinguish between model and modeling, this is useful in some cases, like for instance, cosmology, where `model' usually refers to the cosmological model (e.g., $\Lambda$CDM) and `modeling' refers to the practical application of the model to describe the observables under study. In the case of cosmology, for example, there is a much wider variety of modeling approaches than of models. 
 Assuming a Gaussian likelihood for $\Psi$ (which usually is a good approximation close to its maximum, or in case the central limit theorem applies),  the logarithm of the likelihood given  $M$  depends on $\bm{\theta}$ as 
\begin{equation}
\begin{split}
-2\log \mathcal{L}& \left(\Psi^{\rm d}\vert M,\bm{\theta}\right) =  \\
& =  \sum_{i,j}\left(\Psi_i^{\rm d}-\hat{\Psi}_i(\bm{\theta})\right)\left({\rm Cov}^{-1}(\bm{\theta})\right)_{ij}\left(\Psi_j^{\rm d}-\hat{\Psi}_j(\bm{\theta})\right)^* + \log\left\lvert{\rm Cov} (\bm{\theta}) \right\rvert\, ,
\end{split}
\label{eq:logLkl}
\end{equation}
where $\Psi^{\rm d}$ and $\hat{\Psi}(\bm{\theta})\equiv\hat{\Psi}(\bm{\theta}\lvert M)$ are the data vector and the corresponding theoretical prediction of the observable, respectively, $i$ and $j$ denote vector and matrix indices, and the superscript `*' indicates the complex conjugate operation. The constant terms in $\log\mathcal{L}$, which are not included in Equation~\eqref{eq:logLkl}, do not affect parameter inference, hence we neglect them throughout this work.  In the following, we assume real quantities for $\Psi$ in order to drop the notation referring to the complex conjugate. Nonetheless, we stress that our derivation is equally valid to complex observables.

 Equation~\eqref{eq:logLkl} considers the general case in which the covariance is not fixed but varies as function of the model parameters. For simplicity, in what follows we assume a fixed covariance (i.e., computed for a fiducial model with a specific choice of parameter values). Hence, the second term of the right hand side of Equation~\eqref{eq:logLkl} is constant, and  can be removed from the equation. It is straightforward to extend the methodology to the case of a parameter-dependent covariance matrix following the steps detailed below. Hereinafter we drop the explicit notation for the dependence on the parameters of the model.

\renewcommand{\arraystretch}{1.3}
\begin{table}[]
\resizebox{\textwidth}{!}{
\centering
\begin{tabular}{|c|c||c|c|}
\hline
Symbol & Meaning & Super/sub-script & Meaning \\ \hline\hline
$M$ & Underlying model & $^d$ & Data \\ \hline
$\bm{\theta}$ & Set of model's parameters & $\hat{}$ & Theoretical prediction \\ \hline
$\Psi$ & Observable & $_{i/j}$ & Components of the observable's vector \\ \hline
Cov & Observable's covariance & $^\star$ & Generic point in parameter space \\ \hline
$\mathcal{L}$ & Likelihood & $_\theta$ & In parameter space \\ \hline
$\Delta\bm{\theta}$ & Finite difference in parameter space & $_0$ & Related to the true underlying model \\ \hline
$F$ & Fisher matrix & $^{\rm tr}$ & Actual value in reality \\ \hline
$\bm{\Delta}_{\rm syst}$ & Systematic bias on parameters & $^{\rm fid}$ & Assumed fiducial value \\ \hline
 &  & $^{\rm bf}$ & Best-fit value \\ \hline
 &  & $_{a/b}$ & Components of the parameter's vector \\ \hline
 &  & $^{\rm C}$ & Computed under correct assumptions/approximations \\ \hline
 &  & $^{\rm I}$ & Computed under incorrect assumptions/approximations \\ \hline
  &  & $_{\rm \alpha}$ & Related to an individual likelihood \\ \hline
\end{tabular}}
\caption{Table for reference with the meaning of each of the symbols, subscripts and superscripts used in Section~\ref{sec:shifts}.}
\label{tab:symbols}
\end{table}
\renewcommand{\arraystretch}{1}

For a specific set of  parameters $\bm{\theta}^{\star}$ within a given  model, our target observable is $\hat{\Psi}^\star=\hat{\Psi}(\bm{\theta}^{\star})$. We can approximate the value of the observable under any small variation of the model parameters around $\bm{\theta}^\star$ expanding to linear order as
\begin{equation}
\hat{\Psi}_i(\bm{\theta}^\star+\Delta\bm{\theta}) \approx \hat{\Psi}_i^\star + \left(\bm{\nabla}_\theta\hat{\Psi}_i^\star\right)^T\Delta\bm{\theta},
\label{eq:finidiff}
\end{equation}
where $\bm{\nabla}_\theta$ denotes the gradient operator with respect to the model parameters, the superscript `$T$' denotes the transpose operator, and $\Delta\bm{\theta}$ is a small finite difference in the  parameters space. This approximation is exact when $\hat{\Psi}$ depends linearly on $\bm{\theta}$; otherwise, its accuracy decreases as $\Delta\bm{\theta}$ increases. 

Let us assume that $M_0$, with parameters $\bm{\theta}_0$, is the true underlying model for the phenomenon of interest. Then, $\Psi^{\rm d}$ corresponds to a realization of $M_0$ given $\bm{\theta}_0^{\rm tr}$, with $\hat{\Psi}(\bm{\theta}_0^{\rm tr}) =  \langle\Psi^{\rm d}\rangle $, where the superscript `tr' refers to the  parameter values corresponding to reality and  the brackets $\langle\cdot\rangle$ refer to the ensemble average. \footnote{In certain cases, like in cosmology, it is not always possible to obtain a meaningful ensemble average. In this scenario, we assume that $\bm{\theta}^{\rm tr}$ represent a faithful reproduction of the observed realization.} This is true in practice only if $\Psi^{\rm d}$ is not contaminated with unaccounted-for systematics and an exact theoretical modeling  is used to compute $\hat{\Psi}(\bm{\theta}_0^{\rm tr})$ and its covariance.   
 On the other hand, we can consider a fiducial set of  parameters $\bm{\theta}^{\rm fid}$ within $M$ guessed to be close to the point in parameter space where the likelihood peaks (e.g., inferred from prior or complementary experiments). Note that $M$ and $\bm{\theta}^{\rm fid}$ do not need to be the same model and parameters as $M_0$ and $\bm{\theta}^{\rm tr}_0$; then, in the most general case, we have $\hat{\Psi}(\bm{\theta}^{\rm fid}\lvert M)=\hat{\Psi}^{\rm fid}\neq\Psi^{\rm d}$.  We can expand $\hat{\Psi}^{\rm fid}$ using Equation~\eqref{eq:finidiff}, apply it to Equation~\eqref{eq:logLkl} and maximize the likelihood to obtain the best-fit parameters $\bm{\theta}^{\rm bf}$ for the model $M$. 
  After the maximization, we have
\begin{equation}
\begin{split}
&\sum_{i,j}\left(\bm{\nabla}_\theta\hat{\Psi}_i^{\rm fid}\right)\left({\rm Cov}^{-1} \right)_{ij}\left[\Psi^{\rm d}_j -\hat{\Psi}^{\rm fid}_j-\left(\bm{\nabla}_\theta\hat{\Psi}_j^{\rm fid}\right)^T\Delta\bm{\theta}\right] = 0 \longrightarrow
\\
&\sum_{i,j}\left(\bm{\nabla}_\theta\hat{\Psi}_i^{\rm fid}\right)\left({\rm Cov}^{-1} \right)_{ij}\left(\Psi^{\rm d}_j -\hat{\Psi}^{\rm fid}_j\right) = \sum_{i,j} \left(\bm{\nabla}_\theta\hat{\Psi}_i^{\rm fid}\right)\left({\rm Cov}^{-1} \right)_{ij}\left(\bm{\nabla}_\theta\hat{\Psi}_j^{\rm fid}\right)^T\Delta\bm{\theta},
\end{split}
\label{eq:maxLkl}
\end{equation}
where we have neglected all derivatives of order higher than one. Solving Equation~\eqref{eq:maxLkl} for $\Delta\bm{\theta}$ returns the step in parameter space needed in order to maximize the likelihood:  $\Delta\bm{\theta} \equiv \bm{\theta}^{\rm bf}-\bm{\theta}^{\rm fid}$. The factor multiplying $\Delta\bm{\theta}$ in the right-hand side of Equation~\eqref{eq:maxLkl} can be identified as the Gaussian Fisher information matrix, whose elements  are given by~\cite{Fisher:1935}
\begin{equation}
F_{a,b} = \left\langle \frac{\partial^2\log\mathcal{L}}{\partial\theta_a\partial\theta_b}  \right\rangle = \sum_{i,j} \left(\frac{\partial\hat{\Psi}^{\rm fid}_i}{\partial\theta_a}\right)\left({\rm Cov}^{-1}\right)_{ij}\left(\frac{\partial\hat{\Psi}^{\rm fid}_j}{\partial\theta_b}\right),
\label{eq:Fisher}
\end{equation}
where $a$ and $b$ refer to indices of the parameters vector. 
 Therefore, we can estimate the difference between the best-fit parameters and the fiducial parameters initially assumed as
\begin{equation}
\Delta\bm{\theta} \equiv \bm{\theta}^{\rm bf} - \bm{\theta}^{\rm fid} = F^{-1}\sum_{i,j}\left(\bm{\nabla}_\theta\hat{\Psi}_i^{\rm fid}\right)\left({\rm Cov}^{-1} \right)_{ij}\left(\Psi^{\rm d}_j -\hat{\Psi}^{\rm fid}_j\right).
\label{eq:Deltatheta}
\end{equation}
Of course, if $M$ is a good approximation of $M_0$ and the modeling used to compute $\hat{\Psi}^{\rm fid}$ is accurate, $\bm{\theta}^{\rm bf}$ will be very close to $\bm{\theta}^{\rm tr}_0$.

We can apply the same procedure to a joint analysis of several likelihoods, corresponding to different observables, experiments or independent data sets. However, it is important to note that in general the global best-fit parameters $\bm{\theta}^{\rm bf}$ are different than the best-fit values $\bm{\theta}^{\rm bf}_\alpha$ for each independent likelihood.  Let us consider independent likelihoods, so that the joint likelihood is the product of the individual likelihoods $\mathcal{L}_\alpha$ (i.e., $\log \mathcal{L}=\sum_\alpha \log \mathcal{L}_\alpha$). In this scenario, Equation~\eqref{eq:Deltatheta} is generalized as 
\begin{equation}
\Delta\bm{\theta} = \left(\sum_\alpha F_\alpha\right)^{-1}\left[ \sum_{\alpha,i,j} \left(\bm{\nabla}_\theta\hat{\Psi}_{\alpha;i}^{\rm fid}\right)\left({\rm Cov_\alpha}^{-1} \right)_{ij}\left(\Psi^{\rm d}_{\alpha;j} -\hat{\Psi}^{\rm fid}_{\alpha;j}\right)    \right],
\label{eq:Deltatheta_varL}
\end{equation}
where we denote quantities referred to individual likelihoods with the subscript $\alpha$ (i.e., $\Psi_{\alpha;i}$ refers to the $i$-th element of the observable corresponding to the $\alpha$-th likelihood). If the likelihoods involved in the joint analysis have different nuisance parameters, $F_\alpha$ should be marginalized over the nuisance parameters not common between likelihoods, which will not be included in the parameters vector $\bm{\theta}_\alpha$. Equation~\eqref{eq:Deltatheta_varL} can be straightforwardly generalized to non-independent likelihoods accounting for their covariance in the computation of both the Fisher matrix and the factor in square brackets.

Taking all this into account, we are now ready to compare the performance of a correct and an incorrect modeling of the observable, as well as the effects of assuming an incorrect underlying model. We assume, as before, that $\Psi^{\rm d}$ is drawn from model $M_0$, but this model is unknown. We also consider  two theoretical predictions of the observable, $\hat{\Psi}^{\rm C}$ and $\hat{\Psi}^{\rm I}$, which differ  in the set of assumptions and approximations made for its modeling and in the model assumed: $\hat{\Psi}^{\rm C}$ (correct) uses an accurate and precise modeling assuming a correct  model $M^{\rm C}$, while insufficiently good approximations are implemented, incorrect assumptions are adopted, or an imperfect model $M^{\rm I}$ is used to compute $\hat{\Psi}^{\rm I}$ (incorrect).   
 We can estimate the systematic error $\bm{\Delta}_{\rm syst}$ induced by using $\hat{\Psi}^{\rm I}$ instead of $\hat{\Psi}^{\rm C}$ as 
\begin{equation}
\bm{\Delta}_{\rm syst}  \equiv \bm{\theta}^{\rm bf, I} - \bm{\theta}^{\rm bf, C}, 
\label{eq:systematic}
\end{equation}
where $\bm{\theta}^{\rm bf, I}$ and $\bm{\theta}^{\rm bf,C}$ are obtained using Equation~\eqref{eq:maxLkl} for the correct and incorrect cases introduced above.   

   Our approach to estimate the bias in the inferred parameters can be applied both to real measurements and before they are obtained. As said above, the latter case requires the assumption of a model $M_0$ with  true parameters $\bm{ \theta}^{\rm tr}_0$ as the perfect description of reality; in most forecasts, the assumed fiducial model for the analysis is considered to perfectly describe future observations, hence  $\Psi^{\rm d}=\hat{\Psi}^{\rm fid, C}$. 
 
To make Equation~\eqref{eq:systematic} consistent, if $M^{\rm C}$ and $M^{\rm I}$ are nested models, the parameter space of the model with less parameters should be considered as a hyperplane of the parameter space of the other model, with the values of the extra parameters being kept fixed.  Consider that $\theta_a$ is the extra parameter. In this case, for the model with less parameters we have $\partial\hat{\Psi}/\partial\theta_a=0$. In order to model that this parameter is fixed in the Fisher matrix (and have the same number of parameters in the vectors that enter Equation~\eqref{eq:systematic}), we enforce $F_{*,a}=F_{a,*}\rightarrow\infty$, where the subscript $*$ refer to all indices of the parameters vector. This is equivalent to have a perfect prior on $\theta_a$. 

Equations~\eqref{eq:Deltatheta},~\eqref{eq:Deltatheta_varL}  and~\eqref{eq:systematic} are a generalization of other expressions that have been introduced before for specific cases (see e.g.~\cite{Knox_inhomoreio, Huterer_WLbias, Kim:2004, Huterer_WLbiasPknonlin,  Heavens_modelbias, Taylor_DEshear, Duncan_magnification, Taruya_TNS, Huterer_WLbiasobs, Amara_biasWL, Natarajan_lensingbarfeedback, Pullen_interloperbias, Camera_GRfnlbias, Cardona_magnification, Lorenz_GRbias, Cizmek_magnification, Raccanelli_mnu}). Our expressions, on the contrary, can be applied to the analysis of any given observable, also prior to its measurement. These expressions allows one to estimate the impact  of modeling assumptions and approximations as well as incorrect choices of the underlying model on the inferred parameters. We envision that it will also be useful to single out possible sources of systematic errors affecting new or unexpected findings.

As stated above, the expansion of the observable up to linear order on $\Delta\bm{\theta}$ (Equation~\eqref{eq:finidiff}) is less and less accurate as $\Delta\bm{\theta}$ increases (unless the observable is actually linear on $\Delta\bm{\theta}$). Therefore, in the case the bias introduced on parameter inference is very large, Equation~\eqref{eq:Deltatheta} provides only rough estimations. Of course the estimate of a large bias, even if quantitatively not accurate, is a clear `red flag' for the adoption of the approximation under consideration, hence the approach is still useful. Nevertheless we further quantify the accuracy of the shift in the parameters estimated with Eqs.~\eqref{eq:Deltatheta}, \eqref{eq:Deltatheta_varL}, \eqref{eq:systematic}. In Appendix~\ref{app:shifts_beyond}, we evaluate the performance of the approach outlined in the main text here for a specific case of our case example, the  angular galaxy power spectrum. Moreover, we also discuss how this approach can be  made even more accurate in this appendix. Although we use the linear  expansion discussed in this section in the main body of the text, we  derive the estimation of $\Delta\bm{\theta}$ for an expansion of $\hat\Psi$ up to second order in $\Delta\bm{\theta}$. Finally, we also discuss the estimation of $\Delta\bm{\theta}$ assuming a likelihood of Wishart-distributed variables, as it is the case of the angular galaxy power spectrum. 

\section{Observable and cosmological model considered}
\label{sec:observable}
Although the methodology described above is valid for any measurement, here we focus on its application to cosmology. Given the vast amount of observed data and the complexity of some theoretical calculations, theoretical and numerical approximations are very common. Assessing the reliability of different assumptions might be challenging, but not doing it might have unacceptable consequences. 
 Concretely, we consider the angular galaxy power spectrum as our target observable.  The dramatic improvement in the quality of the observations will require a much better modeling in order to fully exploit coming galaxy surveys. Therefore, approximations that were commonly used in studies about galaxy clustering might  not be accurate anymore. Some of these approximations include, but are not limited to: neglecting  relativistic corrections, the Limber approximation,  
 an incorrect estimation of the covariance matrix, a poor modeling of non-linear clustering and  specific approximations used to model observational effects. For illustrative purposes, in this work we focus on the two first approximations of this list. In Paper I, where the focus is set on the misestimation of the parameter uncertainties, we also study the consequences of neglecting the covariance between different redshifts bins. However, this is expected to have limited impact on the best-fit parameter values; hence, we do not consider this approximation in this study. 

The modeling of the angular power spectrum and of the associated likelihood is discussed with detail in Paper I; here we describe it only briefly, and encourage the reader to refer to Paper I for a full description. The observed galaxy number count perturbations  receive contributions from the intrinsic galaxy overdensities as well as from other effects, such as redshift-space distortions due to peculiar velocities or lensing effects caused by density perturbations along the line of sight. \footnote{Observational  effects (e.g., the observational mask) also modify the observed galaxy overdensity. However, their modeling is extremely case dependent, and will be presented elsewhere. Since we 
 are interested in differential effects, neglecting this is not expected to affect our findings.} 
 All these effects can be modeled in harmonic space introducing several  transfer functions, the  combined effect  of which is given by the total transfer function $\Delta_{\ell}^X(k,z)$ as function of wave number $k$ and redshift $z$, where $\ell$ is the corresponding multipole and $X$ refers to the tracer considered. The explicit form of the contributions from intrinsic clustering, peculiar velocities and relativistic corrections to $\Delta_\ell^X$ can be found in Appendix A of Paper I. $\Delta_\ell^X$ can be restricted to a given redshift bin applying a window function $W(z,z_X,\Delta z_X)$ centered at $z_X$, the width of which is controlled by $\Delta z_X$. For instance, $\Delta z_X$ often refers to the half-width or the standard deviation of a top-hat or  a Gaussian window function, respectively. Accounting for the number density of galaxies per redshift $dN_X/dz$, the transfer  function for a specific redshift bin can be expressed as
\begin{equation}
\Delta^{X,z_X}_{\ell}(k) = \int_{0}^{\infty} dz \frac{dN_X}{dz}W(z,z_X,\Delta z_X)\Delta^X_\ell(k,z),
\label{eq:Delta_l}
\end{equation}
where the integral of $W(z,z_X,\Delta z_X)dN_X/dz$ is equal to unity.  Spectroscopic galaxy surveys provide very precise redshifts  for each galaxy included in the catalog. On the contrary, photometric galaxy surveys sacrifice the precision in the redshift measurements for the sake of observing more galaxies. This uncertainty in the radial position of the galaxies is usually modeled as a smoothing in the radial component of the three-dimensional clustering  in configuration space at small scales. However, the modeling of photometric redshift uncertainties for the angular clustering statistics can be embedded in the choice of window function, given that the radial clustering is projected. Photometric redshifts make impossible to use sharp and narrow redshift bins: the most common choice is to use Gaussian window functions.

With these conventions,  the linear angular galaxy power spectrum for tracers $X$ and $Y$ and redshift bins $z_X$ and $z_Y$, respectively, is given by
\begin{equation}
C^{XY}_\ell(z_X,z_Y) = 4\pi\int\frac{dk}{k} \mathcal{P}_0(k) \Delta^{X,z_X}_{\ell}(k) \Delta^{Y,z_Y}_{\ell}(k),
\label{eq:Cls}
\end{equation}
where $\mathcal{P}_0(k)=k^3P_0(k)/2\pi^2$ is the adimensional,  almost scale-invariant, power-law primordial power spectrum of scalar curvature perturbations.

Galaxies are discrete tracers of the underlying density fluctuations, and therefore their power spectra are affected by shot noise. We assume a Poissonian scale-independent shot noise contribution to be added to the theoretical angular galaxy power spectrum computed in Equation~\eqref{eq:Cls}, hence the total angular power spectrum can be defined as
\begin{equation}
\tilde{C}^{XY}_\ell(z_X,z_Y) = C^{XY}_\ell(z_X,z_Y) + \frac{\delta^K_{z_Xz_Y}\delta^K_{XY}}{dN_X(z_X)/d\Omega},
\label{eq:total_angular_power_spectrum}
\end{equation}
where $dN_X(z_X)/d\Omega$ is the mean number density per steradian for tracer $X$ in the redshift bin centered at $z_X$, and $\delta^K$ is a Kronecker delta. Note that with these assumptions the shot noise term only contributes to the total power spectrum for the auto-power spectrum (i.e., same redshift bin and same tracer). Nonetheless, the shot noise might have non-Poissonian contributions (see e.g.~\cite{Schmittfull_tracers,Ginzburg_halomodel}).  Furthermore, theoretical uncertainties can be added as noise, especially those regarding non-linear scales (see e.g.,~\cite{Montanari_lensingpot}). 

We want to consider all possible combinations of tracers and redshift bins, and denote the number of redshift bins for tracer $X$ and $Y$ with $N_X$ and $N_Y$, respectively. As explained in detail in Paper I, one can consider the angular power spectra or the spherical harmonics coefficients of the galaxy number count fluctuations as the data vector. Depending on the choice, the theoretical $\tilde{C}_\ell$ are used in different manners. For the first option, the power spectra between different redshift bins and tracers at a given multipole are placed in a vector $\bm{C}_\ell$ of size $N_X(N_X+1)/2+N_Y(N_Y+1)/2+N_XN_Y$.  In turn, for the second option, they form matrices $\mathcal{C}_\ell$ of size $(N_X+N_Y)\times (N_X+N_Y)$ which represent the covariance between the spherical harmonic coefficients. Hence, there is a $\bm{C}_\ell$ vector and a $\mathcal{C}_\ell$ matrix for each multipole $\ell$. This choice also determines how the elements of the Fisher matrix are computed~\cite{Tegmark_fisher97}:
\begin{equation}
\begin{split}
F_{ab} = &  \sum_{\ell,i,j} \left(\frac{\partial C_\ell}{\partial\theta_a}\right)_i \left(\mathcal{M}^{-1}_\ell\right)_{ij} \left(\frac{\partial C_\ell}{\partial\theta_b}\right)_j = \\
= & f_{\rm sky}\sum_\ell  \frac{2\ell+1}{2} \sum_{p,q,r,s}\left(\frac{\partial \mathcal{C}_\ell}{\partial\theta_a}\right)_{pq}\left(\mathcal{C}^{-1}_\ell\right)_{qr}\left(\frac{\partial \mathcal{C}_\ell}{\partial\theta_b}\right)_{rs}\left(\mathcal{C}^{-1}_\ell\right)_{sp}\,,
\end{split}
\label{eq:fisher_likelihood_delta}
\end{equation}
where $f_{\rm sky}$ is the fraction of the sky probed by the survey,  $\mathcal{M}_\ell$ is a matrix representing the covariance  between the elements of $\bm{C}_\ell$, indicated by the indices $i$ and $j$. In the second line of Equation~\eqref{eq:fisher_likelihood_delta}, $p$, $q$, $r$ and $s$ refer to the indices of $\mathcal{C}_\ell$; in the sum over these indices, one can recognise the trace of the product of the four matrices involved. We refer the interested reader to Appendix A of Ref.~\cite{hamimeche:cmblikelihood} for further details on the derivation of this expression and the matrix  properties used. 
 The index $i$ of $\bm{C}_\ell$ corresponds to $\tilde{C}_\ell^{(i_1,i_2)}\equiv\tilde{C}_\ell^{XY}(z_X,z_Y)$, where $i_1$ and $i_2$ specify each unique combination of $X$ and $z_X$ and $Y$ and $z_Y$, respectively, of the transfer functions involved in Equation~\eqref{eq:Cls}. 
 Then, each element $i,j$ of the covariance matrix $\mathcal{M}_\ell$ is given by
\begin{equation}
\left(\mathcal{M}_{\ell}\right)_{ij} =  \frac{1}{f_{\rm sky}\left(2\ell+1\right)}\left(\tilde{C}^{(i_1j_1)}_\ell\tilde{C}^{(i_2j_2)}_\ell+\tilde{C}^{(i_1j_2)}_\ell\tilde{C}^{(i_2j_1)}_\ell\right).
\label{eq:Mell_covariance_matrix}
\end{equation}

Now that we have specified the target observable, its covariance, and its Fisher matrix, we can explicitly apply the formalism described in Section~\ref{sec:shifts}. 
 In this case, we can identify the  data and theory vector $\Psi$ as $\bm{C}_\ell$, and its covariance is given by $\mathcal{M}_\ell$. Specifying the general expression in Equation~\eqref{eq:Deltatheta} to the case of the angular galaxy power spectra  we obtain that the difference between the best-fit parameters and the initially assumed fiducial parameters within a general model is 
\begin{equation}
\begin{split}
\Delta\bm{\theta} = & \bm{\theta}^{\rm bf} - \bm{\theta}^{\rm fid} =  F^{-1}\sum_{\ell,i,j} \left(\bm{\nabla}_\theta C_\ell^{\rm fid}\right)_i\left(\mathcal{M}_\ell^{-1}\right)_{ij}\left(C_\ell^{\rm d} - C_\ell^{\rm fid}\right)_j   = \\
= & F^{-1}f_{\rm sky}\sum_\ell \frac{2\ell +1}{2}\sum_{p,q,r,s}\left(\bm{\nabla}_\theta \mathcal{C}^{\rm fid}_\ell\right)_{pq}\left[\left(\mathcal{C}^{\rm fid}_\ell\right)^{-1}\right]_{qr}\left(\mathcal{C}_\ell^{\rm d}-\mathcal{C}_\ell^{\rm fid}\right)_{rs}\left[(\mathcal{C}^{\rm fid}_\ell)^{-1}\right]_{sp}\, .
\end{split}
\label{eq:Delta_theta_Cl}
\end{equation}
  Equation~\eqref{eq:Delta_theta_Cl} can be applied to the correct description of the angular galaxy power spectra and to an approximated one. The substitution of these results in Equation~\eqref{eq:systematic} yields the systematic bias introduced in  parameter inference due to the approximated description, $\bm{\Delta}_{\rm syst}$.

As pointed out in Paper I, Equation~\eqref{eq:fisher_likelihood_delta} is only valid in the case that the same multipole range is used for all redshift bins. This is because both $\mathcal{M}_\ell$ and $\mathcal{C}_\ell$ would be singular otherwise: the matrices corresponding to the  multipoles that are not used in all redshift bins would contain complete rows and columns filled with zeros. Therefore, Equation~\eqref{eq:fisher_likelihood_delta} cannot be applied when the maximum multipole used depends on redshift. This would be the case, for instance, of modeling the redshift dependence of the scales for which non-linear clustering becomes significant.  However, one can consider different likelihoods using a different multipole range for each of them in order to overcome this limitation. Each of these likelihoods includes only the power spectra between the redshift bins that cover the corresponding multipole range. We refer the interested reader to Paper I for more details. Taking this into account, it is straightforward to generalize Equation~\eqref{eq:Delta_theta_Cl} to this case comparing Equations~\eqref{eq:Deltatheta} and Equation~\eqref{eq:Deltatheta_varL}.

\subsection{Cosmological model under study and straw-man survey examples}
\label{subsec:setup}
Given the promising prospects of future galaxy surveys to constrain primordial non-Gaussianities (see e.g.,~\cite{Raccanelli:fNL, Raccanelli2017_PNG, Mueller_fNL, Bernal_EMU, redbook,Karagiannis, euclid}),
 we choose $\Lambda$CDM+$f_{\rm NL}$ to be the cosmological model under study, where $f_{\rm NL}$ parametrizes the amplitude of primordial non-Gaussianity of the local type  controlling the amplitude of the quadractic contributions of a Gaussian random field to the Bardeen potential. The effect of this type of non-Gaussianity on the clustering of halos can be modeled with a strong scale-dependence of the galaxy bias on large scales~\cite{Matarrese_png00,Dalal_png07,Matarrese_png08,Desjacques_png}. Denoting the standard, scale-independent Gaussian galaxy bias as $b_G$ and using the large-scale structure convention for $f_{\rm NL}$, the total galaxy bias is given by
\begin{equation}
b_{\mathrm{tot}}(k,z) = b_G + (b_G-1)f_\mathrm{NL}\delta_\mathrm{crit}\frac{3\Omega_{\rm m}H_0^2}{c^2k^2T(k,z)},
\label{eq:total_galaxy_bias}
\end{equation}
where $\delta_\mathrm{crit}=1.68$ is the critical density related with spherical gravitational collapse in an Einstein-de Sitter cosmology, $\Omega_{\rm m}$ is the matter density parameter today, $H_0$ is the Hubble constant, $c$ is the speed of light, and $T(k,z)$ is the transfer function of matter.\footnote{Detailed comparison with N-body simulations indicate that there might be a correction factor of order unity to Equation~\eqref{eq:total_galaxy_bias} (see e.g.,~\cite{Margot, WagnerNG}), which for simplicity we omit here. Hence, $f_{\rm NL}$ in Equation~\eqref{eq:total_galaxy_bias} should be considered as an effective primordial non-Gaussianity parameter of about the same magnitude of the true underlying $f_{\rm NL}$.}

Including the Gaussian galaxy bias of each tracer as a model parameter, the set of parameters considered in this work to model $\Lambda$CDM$+f_{\rm NL}$ is $\bm{\theta} = \left\lbrace h, \omega_\mathrm{b}, \omega_\mathrm{cdm}, n_\mathrm{s}, b_G^X, b_G^Y, f_\mathrm{NL}  \right\rbrace$,\footnote{We do not consider the amplitude of $\mathcal{P}_0$ as a free parameter in our analysis because it is almost  completely degenerate with $b_G^X$ and $b_G^Y$, although we are aware that it should be included in an analysis of real observations.} where $\omega_{\rm b}$ and  $\omega_{\rm cdm}$ are the physical densities of baryons and cold dark matter, respectively, $n_s$ is the spectral index of $\mathcal{P}_0(k)$, and  $b_G^X$ ($b_G^Y$) is the Gaussian galaxy bias for the tracer $X$ ($Y$).  Note that for analyses with only one tracer, $b_G^X$ and $b_G^Y$ become simply $b_G$.  

As done in Paper I, we consider three straw-man galaxy surveys to study if the systematic biases depend qualitatively on the survey parameters, as well as to study the multi-tracer case. First, we consider a survey with galaxies uniformly  distributed in redshift and a galaxy density per unit redshift and square degree $d^2N_\mathrm{g}/dzd\Omega = 1070\ \mathrm{gal/deg}^2$. We also consider two other more realistic surveys, inspired by the galaxy redshift distribution expected for Euclid~\cite{euclid} (in a pessimistic scenario) and SPHEREx~\cite{spherex},  which we approximate by
\begin{equation}
\frac{d^2N}{dzd\Omega} = A\left(\frac{z}{z_0}\right)^\alpha e^{-\left(z/z_0\right)^{1.5}}\, {\rm gal/deg^2}\, ,
\label{eq:dNdzdOmega}
\end{equation}
with $A^{\rm Eu-l} = 2.4\times 10^3$, $z_0^{\rm Eu-l}=0.54$, and $\alpha^{\rm Eu-l}=4.0$ for Euclid; and $A^{\rm SP-l}=2.93\times 10^4$, $z_0^{\rm SP-l}=0.53$ and $\alpha^{\rm SP-l}=1.1$. We denote these two galaxy distributions as Euclid-like and SPHEREx-like, respectively, with corresponding related superscripts `Eu-l' and `SP-l'.  We normalize the number density of the uniformly sampled survey to contain the same number of galaxies as our Euclid-like survey.  

We consider  a full redshift range $0.1\leq z\leq 2.1(2.2)$, split into five redshift bins centered at $z = 0.3,\,0.7,\,1.1,\,1.5$, and $1.9$ for the three galaxy surveys considered. All redshift bins have a top-hat window function with half-widths of 0.2 or 0.3 (so that the redshift window functions of adjacent redshift bins overlap only in the second case); we refer to these two cases as `non-overlap' and `overlap' cases, respectively.   The overlapping case with top-hat window functions can be understood as a worst-case scenario for limited photometric redshift uncertainties. While Gaussian window functions overlap with each other, it is the non-overlapping intervals which contribute the most to the angular power spectrum (see Equation~\eqref{eq:Delta_l}). On the other hand, top-hat window functions give equal weight to all redshifts  within the  bin. Therefore, the results for a photometric survey would correspond to a  very survey-dependent intermediate case between our overlapping and non-overlapping cases. 
 In all cases considered, as in Paper I, we
consider $f_{\rm sky}= 1$. It is straightforward to rescale our results for different values of $f_{\rm sky}$. 

Finally, we need to set the multipole range that will be used. We explore two different scenarios. First, we consider $\ell\in \left[ 2,\ell_{\rm max}(z)\right]$ unless otherwise stated, where $\ell_{\rm max}$ is the multipole corresponding to the smallest  scale $k_{\rm max}$ for which non-linear clustering can be neglected: $\ell_{\rm max}=k_{\rm max}\chi(z)$, where $\chi(z)$ is the comoving distance to the mean redshift of the redshift bin of interest. We assume $k_{\rm max}=R_{\rm max}^{-1}$, where $R_{\rm max}$ is the radius of a top-hat window in configuration space for which the variance of the matter fluctuations within a given radius in configuration is unity: $\sigma_{\rm m}(R_{\rm max})=1$.  We use $\ell_{\rm max}=\lbrace 180,550,1100,1900,3000\rbrace$ for $z=\lbrace 0.3,0.7,1.1,1.5,1.9 \rbrace$, respectively. In addition, we consider a conservative case in which $\ell_{\rm max}=200$ for all redshift bins.

\section{Systematic bias induced by different approximations}
\label{sec:results}
  The results shown in this section aim to be an example of the performance of the methodology described in Section~\ref{sec:shifts}, but also to act as a warning for future measurements. For illustration purposes, we choose to apply the methodology to a case where the bias introduced in the best-fit parameters arises only from incorrect modeling. Therefore, in what follows we consider that  the only model under study matches the true cosmology, so that $M^{\rm C}=M_0$, while $M^{\rm I}$ refers to the same underlying model but when an incorrect modeling of the observable is used. Furthermore, we take $\bm{\theta}^{\rm fid,I}=\bm{\theta}^{\rm fid,C}=\bm{\theta}_0^{\rm tr}$.   
  This means that the data drawn from $M_0$ is equal to the prediction for $M^{\rm C}$ using the correct modeling: $C_\ell^{\rm d}\equiv\hat{C}_\ell^{\rm C}(\bm{\theta}^{\rm fid,C}\lvert M^{\rm C})$.  We choose $M_0$ to be $\Lambda$CDM+$f_{\rm NL}$, with parameter values $\bm{\theta}_0^{\rm tr}$: $h=0.6727$, $\omega_{\rm b}=0.02225$, $\omega_{\rm cdm} = 0.1198$, $n_s=0.9645$, $f_{\rm NL}=0$,  an amplitude of the primordial power spectrum $\ln 10^{10}A_s = 3.0940$, and we consider three massive degenerate neutrinos with mass $m_\nu=0.02$ eV each.  We assume scale- and redshift-independent Gaussian galaxy bias for the sake of simplicity: $b_G^{\rm unif}=b_G^{\rm Eu-l}=2$ for the uniform and Euclid-like surveys and $b_G^{\rm SPx}=1.4$ for the SPHEREx-like survey; this assumption does not change the qualitative results of the examples under study.

The modeling we use to compute $C_\ell^{\rm C}$ and $C_\ell^{\rm d}$ includes  relativistic corrections and  redshift-space distortions due to peculiar velocities and lensing magnification, and does not use the Limber approximation.  In turn, $C_\ell^{\rm I}$ differs from them in one aspect of the modeling: in Section~\ref{sec:shift_lensing} $C_\ell^{\rm I}$ does not include the contribution from lensing magnification, while in Section~\ref{sec:shift_limber} $C_\ell^{\rm I}$ uses the Limber approximation.  Both cases are explored using a multi-tracer analysis of the angular galaxy power spectra in Section~\ref{sec:shift_multi}. 

 In all cases considered below, the estimated systematic biases correspond to the full vector of the shift in  the multidimensional parameter space.  Afterwards, we assess the significance of the estimated biases in the marginalized constraints on each of the parameters by comparing them with the 68\% confidence level of their respective marginalized uncertainties, obtained using the same approximations.  As a general rule of thumb,  the parameter $\theta_a$ would be considered significantly biased if the corresponding component of $\bm{\Delta}_{\rm syst}$, $\Delta_{{\rm syst},a}$, is larger than the 68\% confidence level marginalized uncertainty in the inference of $\theta_a$: $\Delta_{{\rm syst},a}/\sigma_{\theta_a} \gtrsim 1$.  We refer the interested reader to Paper I for a detailed discussion on the effects of approximations on the parameter uncertainties.

\subsection{Ignoring lensing magnification}
\label{sec:shift_lensing}
Matter density perturbations along the line of sight affect how we observe the galaxy density distribution~\cite{Yoo_GR09}. 
 Therefore, the observed galaxy number count perturbations are determined by the  intrinsic clustering, redshift-space distortions due to peculiar velocities and relativistic corrections. These corrections can be separated into contributions from lensing magnification, doppler, and gravitational potential effects such as time-delays and integrated Sachs-Wolfe effect~\cite{Yoo_GR10, Bonvin_obsLSS, Challinor_obsLSS, Jeong_GR12, Bertacca_xi3dGR, Raccanelli_GRCl}. 

Lensing magnification is a subdominant contribution to the observed galaxy overdensities except when the redshift separation between bins is large enough so that the correlation due to intrinsic clustering is negligible. However, it is normally the largest relativistic correction, especially when cross-correlating two different redshift bins using the angular power spectrum~\cite{Raccanelli_GRint}. 
Some arguments against including lensing magnification  include its subdominant relative contribution to the galaxy power spectrum at the scales explored so far,  the large computational expenses required for its calculation and the difficulties in obtaining an accurate determination of the magnification bias parameter (see e.g.,~\cite{Hildebrandt_magnification_biases}). 
  Nonetheless, the magnification contribution contains  cosmological information complementary to weak-lensing shear~\cite{VanWaerbeke_magnification,Casaponsa_magnification, Duncan_magnification}. 

To understand and model the effect of lensing magnification consider that the gravitational lensing contribution to the galaxy overdensity consists of two competing effects: on the one hand, it stretches the volume behind the lens; on the other, it magnifies individual sources and promotes faint galaxies above the magnitude limit of the survey~\cite{turner:magnificationbias}. This changes the observed galaxy number density $n_{\rm obs}$ in a flux-limited survey: 
\begin{equation}
n_\mathrm{obs} = n\left[ 1 + (5s-2)\kappa \right],
\end{equation}
where $n$ is the intrinsic galaxy number density, $s$ is the magnification bias parameter, and $\kappa$ is the convergence~\cite{Bartelmann_mag01}. Note that $s=0.4$ corresponds to a vanishing contribution from lensing magnification. Since the magnification bias parameter depends on the tracer used, we distinguish between $s^X$ and $s^Y$.  We do not consider $s$ as a nuisance parameter (as should be done in a more quantitative analysis of actual observations); instead, we study several cases with  different constant values of $s$ in order to study the dependence of the systematic biases on this parameter.  

\begin{figure}
\centering
\includegraphics[height=0.6cm]{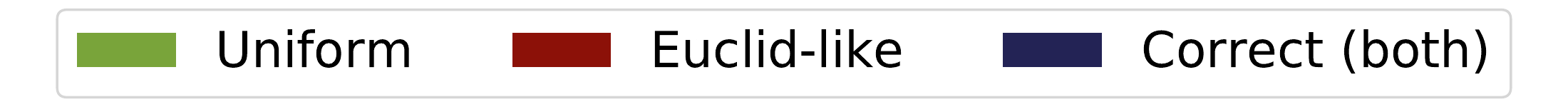}
\includegraphics[height=0.6cm]{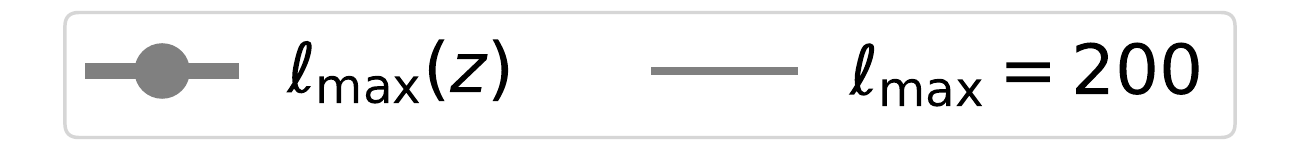}
\includegraphics[width=1\linewidth]{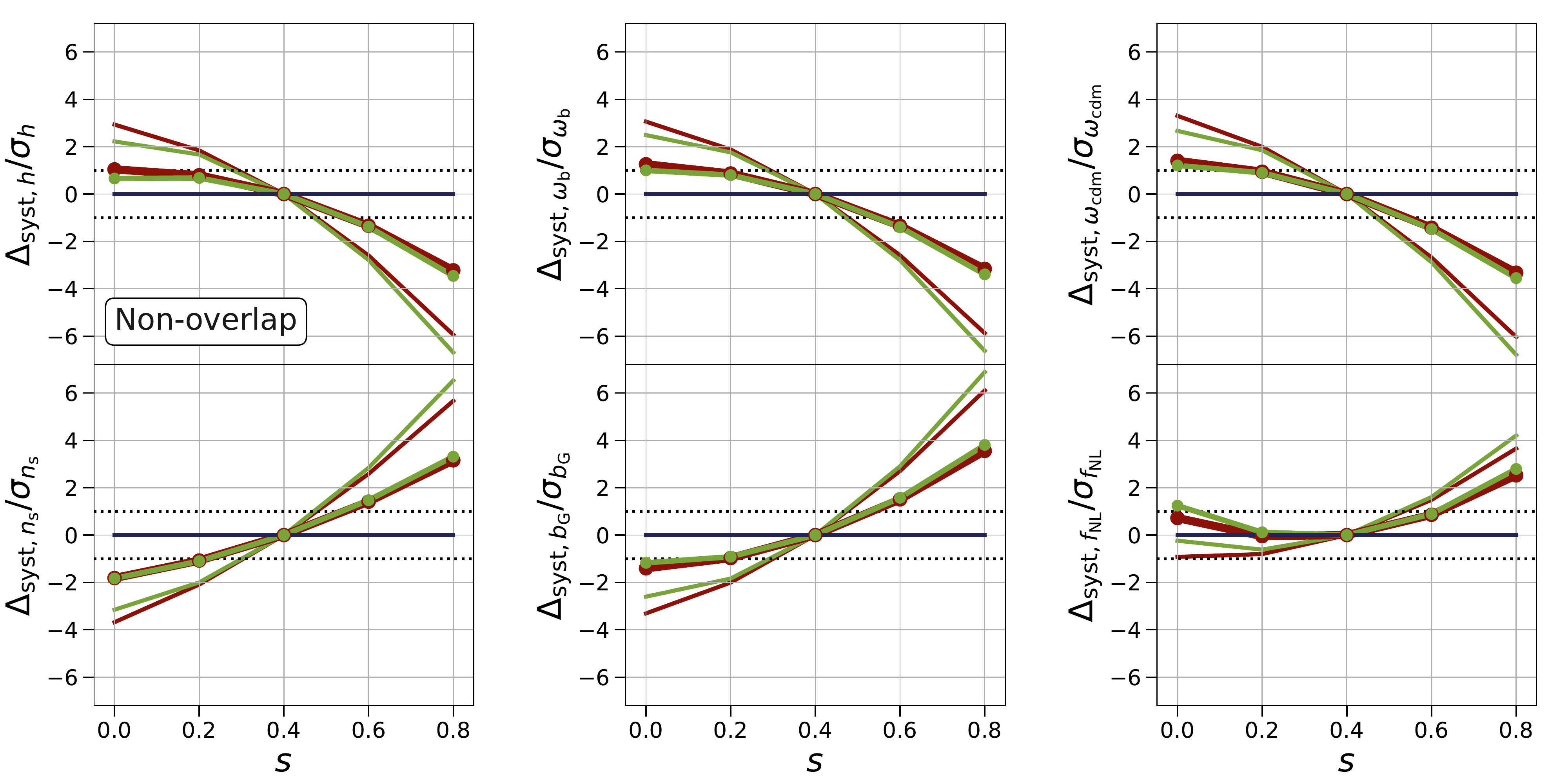}
\includegraphics[width=1\linewidth]{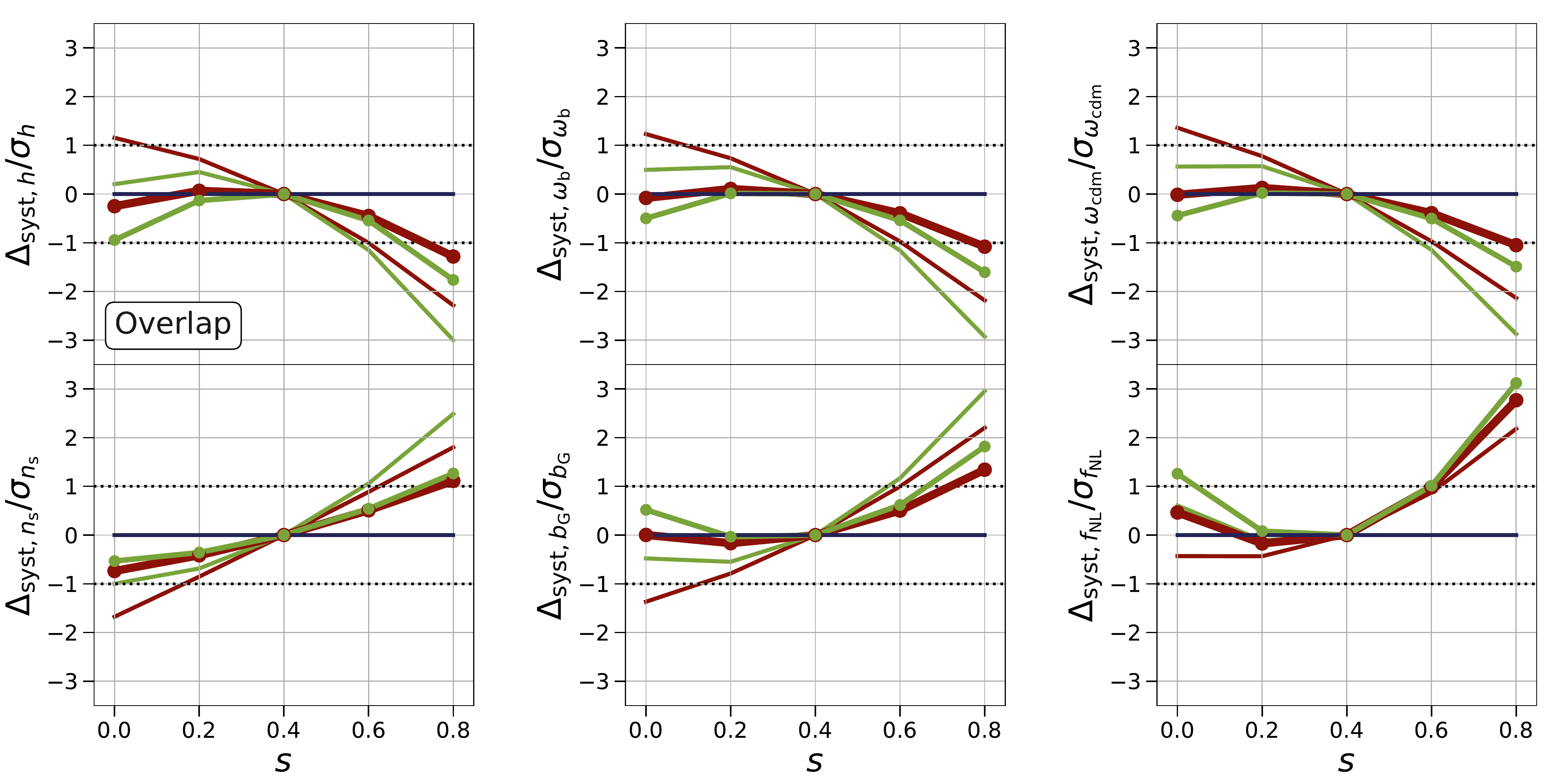}
\caption{Ratio of the estimated bias in the model parameters over the forecasted 68\% confidence level marginalized constraints  for cases when the lensing magnification is included (blue) and not as a function of the assumed magnification bias parameter. We show results for uniform (green) and Euclid-like (red) galaxy redshift distributions, and with non-overlapping (top) and overlapping (bottom) redshift bins; note the change of scale in the $y$-axis between them. In all cases, the case for $\ell_{\rm max}(z)$ is shown with wide solid lines with circle markers, while thin solid lines without markers correspond to the case with constant $\ell_{\rm max}=200$. Dotted lines mark $\lvert\Delta_{{\rm syst},a}/\sigma_{\theta_a}\lvert=1$. }
\label{fig:relshifts_magnification}
\end{figure}

We show the significance of the estimated biases when lensing magnification is neglected as a function of the magnification bias parameter in  Figure~\ref{fig:relshifts_magnification}. Considering only single-tracer analyses, we show results for the uniform and Euclid-like galaxy surveys, and normalize the estimated biases with the forecasted 68\% confidence level marginalized constraints in order to show the bias significance. We find that the size of the systematic bias grows as $\vert s-0.4\vert$ increases, and that it grows faster for magnified populations (i.e., $s>0.4$), than for de-magnified.  Results for different redshift distributions are qualitatively very similar. The only difference is that, for $\ell_{\rm max}=200$,  the estimated biases are larger for the uniform survey than for the Euclid-like survey when the redshift bins overlap, and viceversa when the redshift bins do not overlap.  While for the Euclid-like survey the biases are always larger when the redshift bin do not  overlap, this is only true for the uniform survey when $s>0.4$. 

In general, the significance of the bias is much larger for the case in which $\ell_{\rm max}=200$ than when $\ell_{\rm max}$ varies with redshift (reaching higher values). This may be counter-intuitive, since the signal-to-noise ratio of the magnification contribution increases at smaller scales. However, the higher-$\ell$ part is computed only for the highest-z bins (see Section~\ref{subsec:setup}), so that the relative contribution of lensing magnification is smaller at these scales (remember that lensing magnification dominates the angular power spectra for very separated redshift bins). Moreover, the constraints on the cosmological parameters also improve for increasing $\ell_{\rm max}$, and $\Delta\bm{\theta}\propto F^{-1}$. The reduction of the significance of the bias with higher $\ell_{\rm max}$ can be understood as the information beyond lensing magnification encoded in the angular galaxy power spectra having more weight in the final parameters constraints, with respect to the $\ell_{\rm max}=200$ case.

\begin{figure}
\centering
\includegraphics[width=\linewidth]{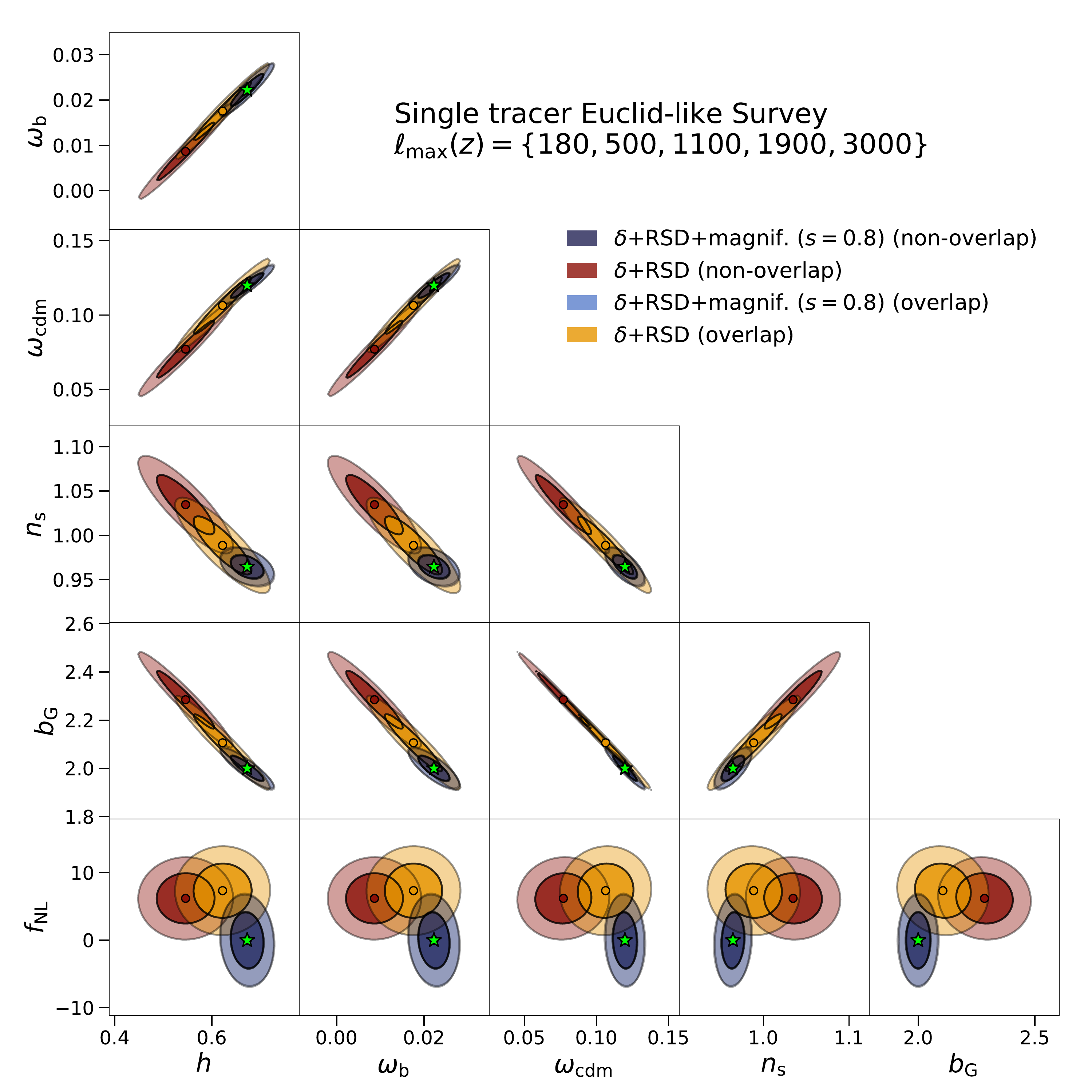}
\caption{68\% and 95\% confidence level estimated marginalized constraints on the model parameters for a Euclid-like survey with a single tracer with $s=0.8$ using $\ell_{\rm max}(z)$. We compare the results including cosmic magnification (blue) and without including it (red), assuming that the fiducial cosmology assumed coincides with the actual one, marked with a star. Dark colors (blue and red) refer to the case with non-overlapping redshift bins, while light colors (light blue, orange), refer to the case with overlapping redshift bins.}
\label{fig:corner_lensing}
\end{figure}

Assessing the significance of the biases exploring only one-dimensional marginalized parameter constraints might be misleading. The bias might be much more significant for systematic shifts in perpendicular directions  to parameter degeneracies than what would be inferred from one-dimensional projections.  We show forecasted two-dimensional marginalized constraints for all parameter combinations from a single-tracer Euclid-like survey in the $\ell_{\rm max}(z)$ case including the estimated biases with respect to the fiducial cosmology in Figure~\ref{fig:corner_lensing}.  We compare the case including magnification (with $s=0.8$) and the case without including magnification (shown in blue and red, respectively). We show results for the non-overlap (overlap) redshift bin configuration in dark (light) colors. As found in Paper I, the parameter degeneracies obtained neglecting lensing magnification are not necessarily the same as for the correct analysis. Furthermore, we find   the systematic bias to be aligned with the degeneracies of the incorrectly estimated constraints.  Whether the alignment is general or depends on the observable is  beyond the scope of this work. As expected from Figure~\ref{fig:relshifts_magnification}, the biases shown in Figure~\ref{fig:corner_lensing} are larger for the case with non-overlapping redshift bins. Finally, while the correct uncertainties do not depend on whether the redshift bins overlap or not, the resulting incorrect confidence level regions  (i.e., neglecting magnification) are slightly larger when the redshift bins overlap (which further reduces the significance of the bias). The results using $\ell_{\rm max}=200$ are qualitatively similar (with a higher significance of the bias, as shown in Figure~\ref{fig:relshifts_magnification}).

\subsection{Using the Limber approximation}
\label{sec:shift_limber}
\begin{figure}
\centering
\includegraphics[width=\linewidth]{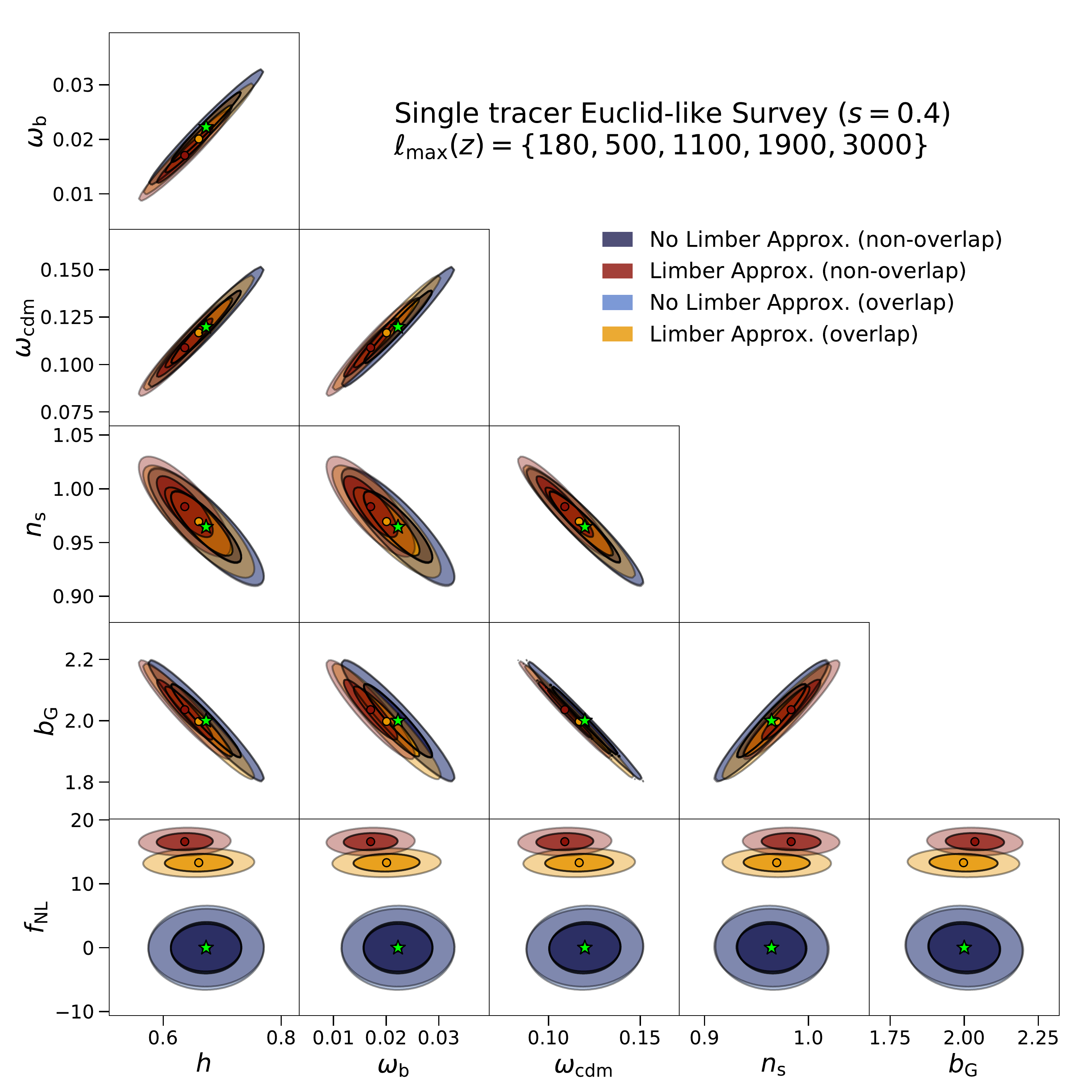}
\caption{Same as Figure~\ref{fig:corner_lensing}, but comparing  the results with and without using the Limber approximation and assuming $s=0.4$. }
\label{fig:corner_limber}
\end{figure}

As shown in Equation~\eqref{eq:Cls}, the angular power spectrum is the projection along the line-of-sight of the spatial, three-dimensional power spectrum. The transfer functions that drive this projection include spherical Bessel functions. Therefore, integration over two spherical Bessel functions is needed  to compute $C_\ell$ for each multipole. Given the oscillatory nature of the spherical Bessel functions, it is very computationally expensive to ensure the convergence of these integrals, which slows down the calculation of $C_\ell$.  The Limber approximation~\cite{limber:approximation,kaiser:limberapproximation,loverde:extendedlimber} aims to alleviate this problem by approximating the spherical Bessel functions as Dirac delta functions. However, the transformation of low-order spherical Bessel functions (i.e., low $\ell$) into  Dirac delta functions is not accurate, which means that the Limber approximation breaks down for $C_\ell$ at large scales. Using the Limber approximation to compute the galaxy power spectrum may introduce significant biases in cosmological parameters, even when combined with galaxy weak lensing observations, for which the Limber approximation is accurate enough (see e.g.~\cite{Fang_limber}).

We show two-dimensional forecasted marginalized constraints on the cosmological parameters from the angular power spectrum of a single-tracer Euclid-like galaxy survey in Figure~\ref{fig:corner_limber}. 
 We show results using $\ell_{\rm max}(z)$ both with and without the Limber approximation (red and blue, respectively), and show the results with overlapping redshift bins in lighter colors. We assume $s=0.4$ in order to avoid  contributions from lensing magnification. As noted in Paper I, parameter uncertainties are underestimated when the Limber approximation is used, especially for non-overlapping redshift bins. This makes the (generally) small  shifts in the best-fit values more significant when compared to the small forecasted errors. The bias is especially worrisome for $f_{\rm NL}$, the only example where the shift in the best fit is very large. This is because the signature of local primordial non-Gaussianities manifests at large scales, which is where the Limber approximation breaks down. The estimated systematic bias in $f_{\rm NL}$ is  $\gtrsim 18(14)\sigma$  for the case with non-overlapping (overlapping) redshift bins. This result indicates that using the Limber approximation at large scales potentially leads to a false positive detection of primordial non-Gaussianity.   In general,  biases are smaller when the redshift bins overlap, as it was the case for the non-inclusion of lensing magnification of the signal. The results using $\ell_{\rm max}=200$ are qualitatively similar, with the weaker constraints (due to the use of a narrower multipole range) but with a comparable significance of the bias due to the use of the Limber approximation. This is because the Limber approximation breaks down at low $\ell$, which is a regime covered in both cases.  

\subsection{Multi-tracer galaxy power spectrum}
\label{sec:shift_multi}
One of the most exciting prospects offered by the  next-generation galaxy surveys is the possibility to perform multi-tracer analyses~ \cite{Seljak_fNLmultitracer,McDonald_multitracer}. Instead of using all galaxies as a single tracer, using various tracers of the underlying density field at the same time and fully accounting  for their cross-correlations, provides additional information. The gain comes from probing the  same volume  more than once, each time with a different galaxy bias, which reduces the cosmic variance for quantities that are related to the galaxy bias. Besides tightening the constraints on all cosmological parameters, especially those related with the galaxy bias,  the use of multi-tracer approaches are expected to be especially effective to constrain physics that affect the large scales, such as local primordial non-Gaussianities.  

Given the reduced cosmic variance, systematic errors may produce more significant biases in  parameter inference than for single-tracer studies. We apply the methodology described in Section~\ref{sec:shifts} to multi-tracer analyses of the angular galaxy clustering  in order to asses the impact of the approximations discussed above. 
 We use \texttt{Multi\_CLASS} to compute the angular cross-power spectrum of two different tracers (with their own redshift distribution, galaxy bias and magnification bias parameter). We consider two different galaxy surveys (or galaxy populations observed by the same survey) following a Euclid-like and a SPHEREx-like redshift distribution and with different galaxy bias as specified in Section~\ref{subsec:setup}. 

\begin{figure}
\centering
\includegraphics[width=\linewidth]{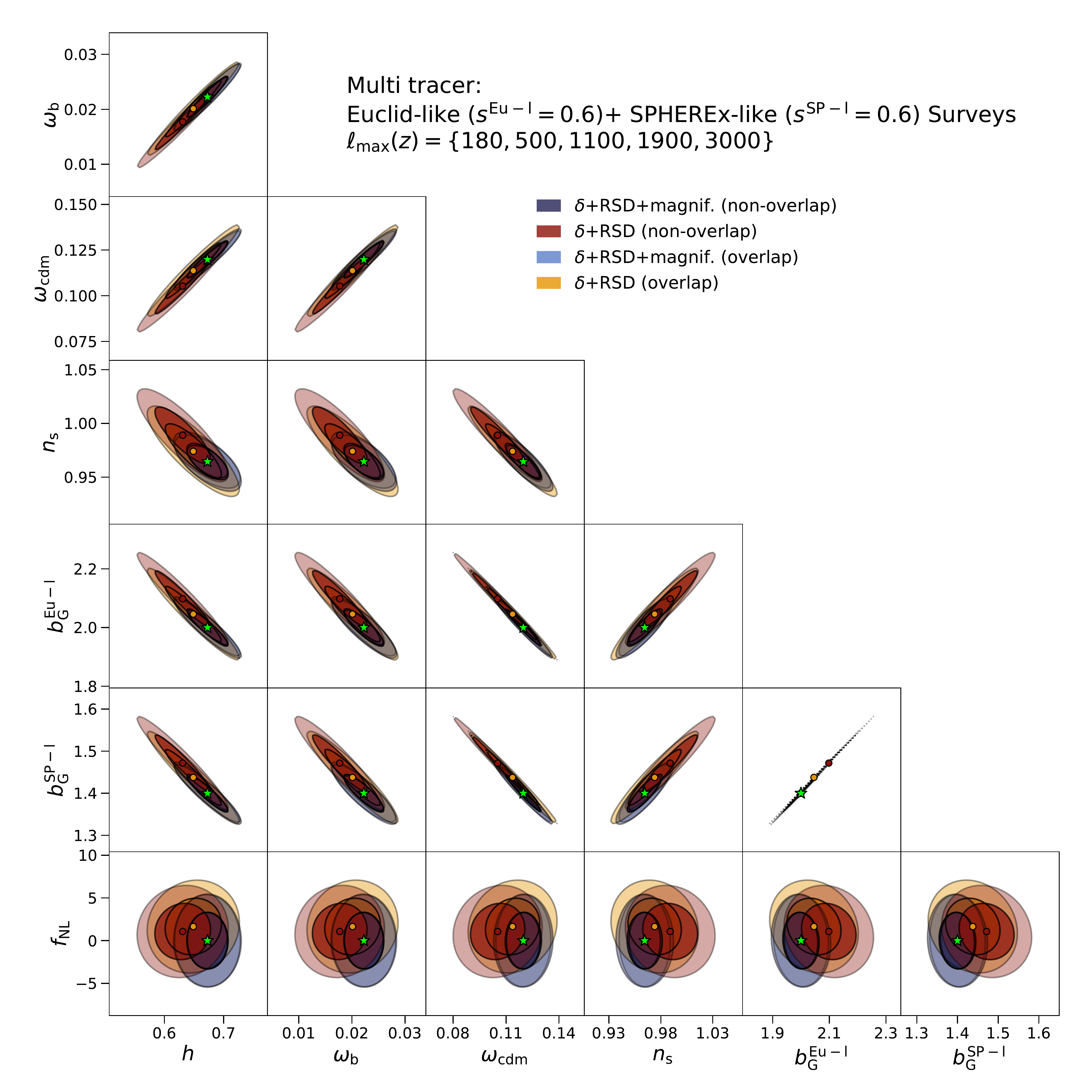}
\caption{Same as Figure~\ref{fig:corner_lensing}, but for a multi-tracer analysis of the Euclid-like and SPHEREx-like surveys considered in this work, assuming $s=0.6$ for both surveys. }
\label{fig:multi_corner_lensing}
\end{figure}

\begin{figure}
\centering
\includegraphics[width=\linewidth]{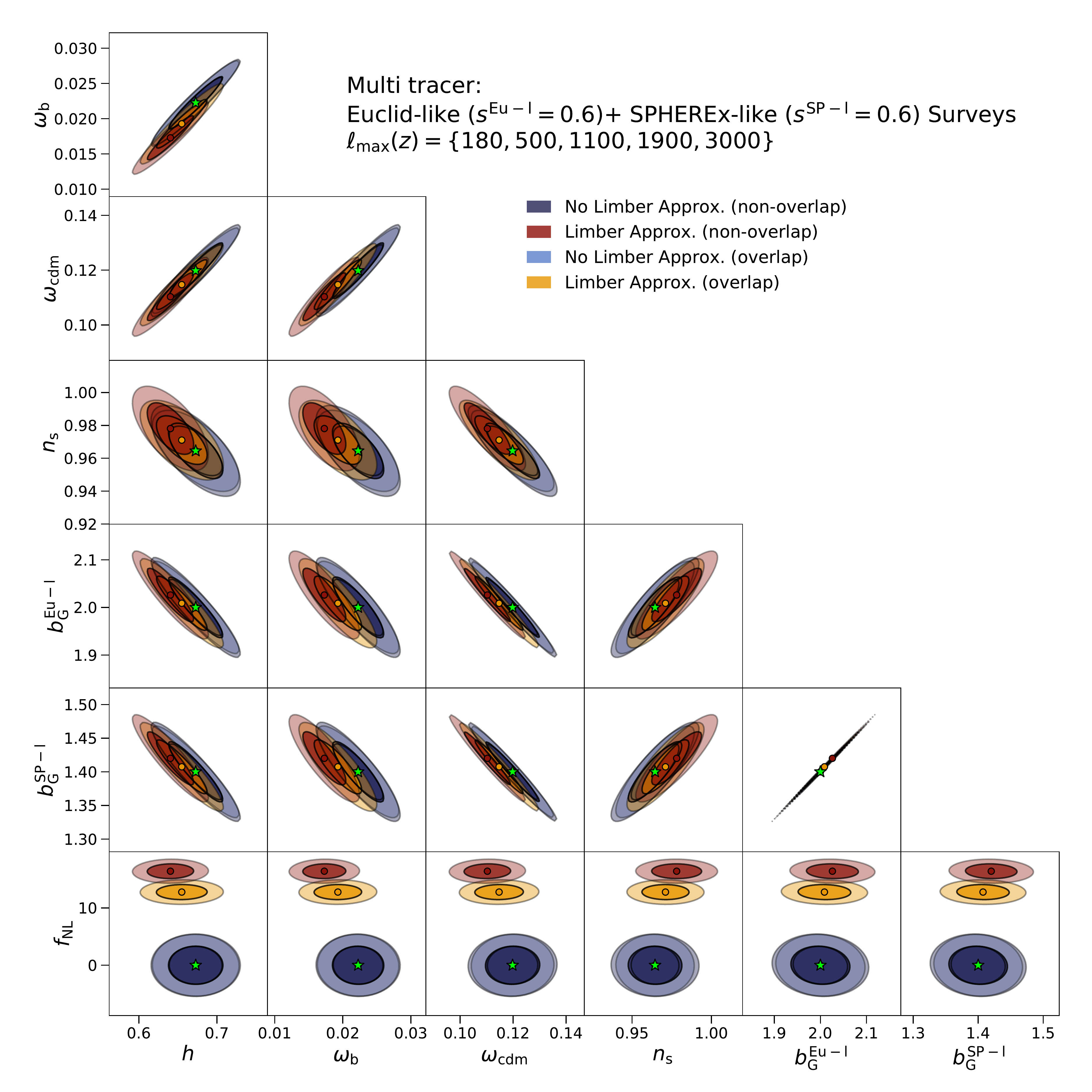}
\caption{Same as Figure~\ref{fig:corner_limber}, but for a multi-tracer analysis of the Euclid-like and SPHEREx-like surveys considered in this work, assuming $s=0.6$  for both surveys.}
\label{fig:multi_corner_limber}
\end{figure}

We show forecasted two-dimensional marginalized constraints of the cosmological parameters under study with and without modeling lensing magnification and using $\ell_{\rm max}(z)$ in the signal and covariance in Figure~\ref{fig:multi_corner_lensing}. For illustrative purposes, we focus just on the case with magnification bias parameters $s^{\rm Eu-l}=s^{\rm SP-l}=0.6$ for the Euclid-like and SPHEREx-like surveys. The figure also includes the estimated bias in the best-fit parameters. We can appreciate that, as reported in Paper I, ignoring lensing magnification overestimates the uncertainties in the parameters, except for $f_{\rm NL}$ (and $\omega_{\rm b}$, for which there is practically no effect). Moreover, there are still significant biases ($\sim 1-2\sigma$) in all  parameters when the redshift bins do not overlap (shown in darker colors). Surprisingly, the systematic bias is much less significant ($\lesssim 1\sigma$) when the redshift bins do  overlap. As in Figure~\ref{fig:corner_lensing}, the estimated bias is aligned with the degeneracy between the parameters, which almost does not change whether lensing magnification is included or not. The alignment between the bias and the parameter degeneracies is most likely not generic, but very population selection (and thus, survey) dependent. With a slightly different set up (or different fiducial $s$ for the two populations) this alignment may not hold. Such scenario may lead to great impact in final results: a change in the degeneracies would greatly exacerbate the effect of the systematic bias introduced by approximations when combining the angular galaxy power spectra with other cosmological probes.

Similarly to the single-tracer case, the estimated bias found when using $\ell_{\rm max}=200$ is larger than using $\ell_{\rm max}(z)$, but the degeneracies and direction of the shifts remain unchanged. The increase of the significance of the bias using  $\ell_{\rm max}=200$ instead of $\ell_{\rm max}(z)$ is smaller than for the single-tracer case. This is because the multi-tracer approach reduces the cosmic variance, so that the relative contribution to the constraints from large scales in the $\ell_{\rm max}(z)$ case is higher than in the single-tracer case and the effect of using $\ell_{\rm max}=200$ is smaller.

Figure~\ref{fig:multi_corner_limber} shows the analogous results for using or not the Limber approximation (and considering lensing magnification in both cases). The uncertainties of the parameters are underestimated at approximately the same level as for the single-tracer case, but the estimated biases are larger in this case. Note also that in some cases the biases is not aligned with the parameter degeneracies.  As for the Euclid-like only analysis shown in the previous section, the biases are smaller for overlapping bins (lighter colors) than for non-overlapping bins, because when the redshift bins overlap biases are smaller and incorrect uncertainties are larger.  In this case $f_{\rm NL}$ is again  by far the most affected parameter, with biases of $19 \sigma\, (14\sigma)$ for non-overlapping (overlapping) redshift bins. In this case, as for the single-tracer case, the results using $\ell_{\rm max}=200$ are very similar to those shown in Figure~\ref{fig:multi_corner_limber}.

\begin{figure}
\centering
\includegraphics[height=0.6cm]{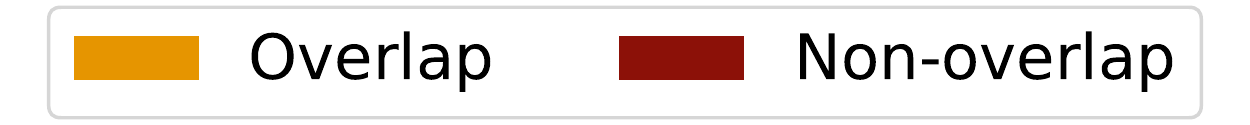}\\
\includegraphics[height=0.6cm]{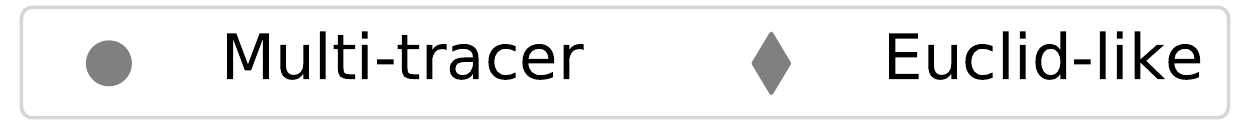}
\includegraphics[height=0.6cm]{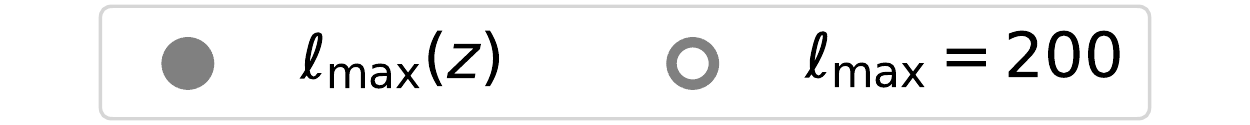}
\includegraphics[width=\linewidth]{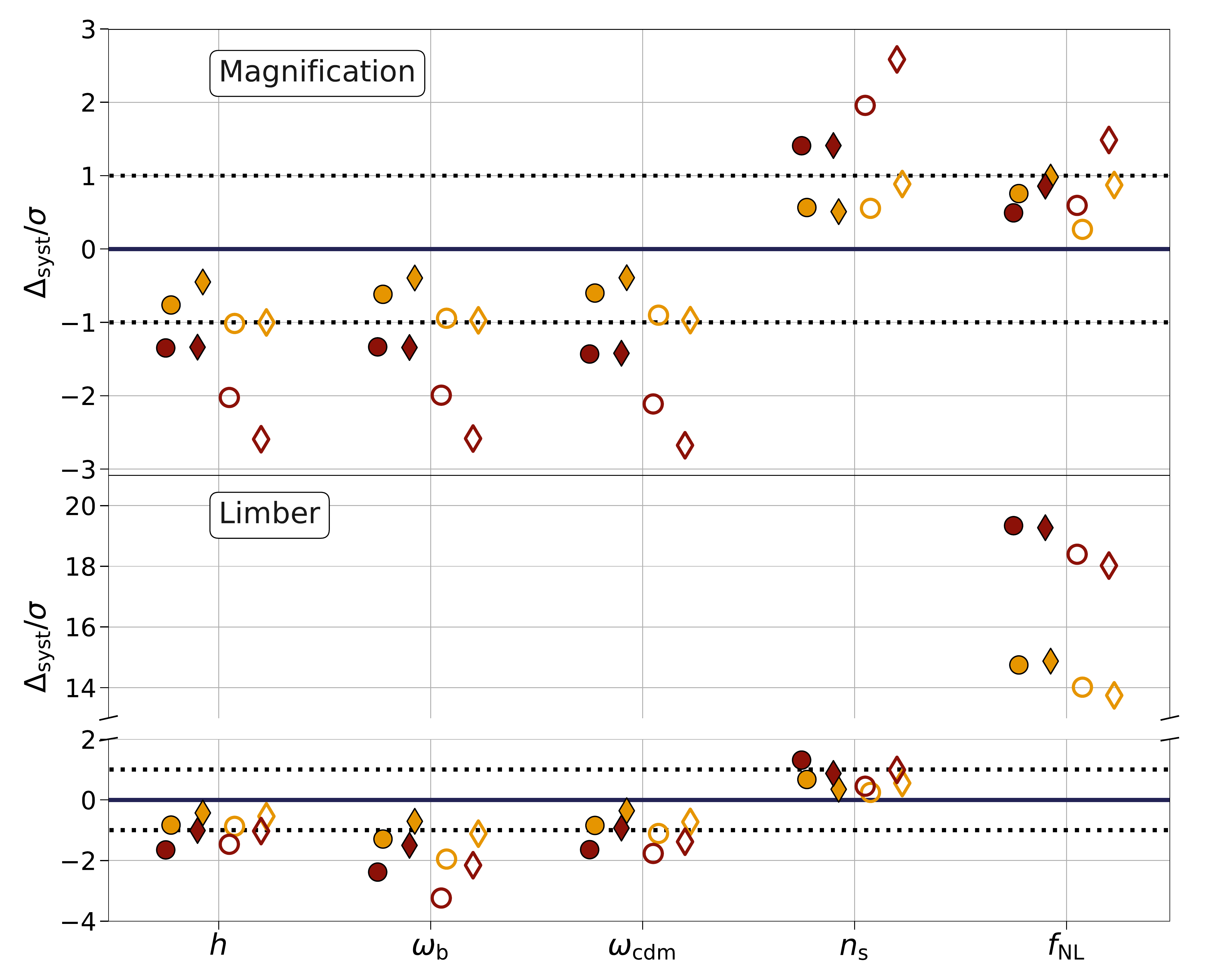}
\caption{Ratio of the estimated bias in the cosmological parameters over the forecasted 68\% confidence level marginalized constraints  for cases when the lensing magnification is not modeled (upper panel) and the Limber approximation is used (bottom panel).  We show results considering an Euclid-like survey (diamonds), and its combination with a SPHEREx-like survey performing a multi-tracer analysis (circles), both for overlapping (orange) and non-overlapping (red) redshift bins. In all cases we consider $s=0.6$ for both galaxy populations. Filled (hollow) markers refer to results using $\ell_{\rm max}(z)$ ($\ell_{\rm max}=200$), respectively. Note the different scale in the $y$-axis in each panel, and that the $y$-axis in the right panel is broken. Dotted lines mark $\lvert\Delta_{{\rm syst},a}/\sigma_{\theta_a}\lvert=1$.}
\label{fig:multi_single}
\end{figure}

Finally, we can compare the estimated marginalized biases for the Euclid-like only analyses and for the multi-tracer analysis combining the Euclid-like and the SPHERE-like surveys. We show the ratio of the estimated biases over the forecasted 68\% confidence level marginalized constraints for both cases (using both $\ell_{\rm max}(z)$ and $\ell_{\rm max}=200$) in Figure~\ref{fig:multi_single}. We show results both for overlapping and non-overlapping bins, and using separately the two approximations considered in this work: neglecting lensing magnification (left panel), and using the Limber approximation (right panel). Figure~\ref{fig:multi_single} shows in clearer way the comparison between the single-tracer and multi-tracer cases and using different criteria for $\ell_{\rm max}$  discussed above. When lensing magnification is not included the biases are approximately the same for both single- and multi-tracer analyses if $\ell_{\rm max}(z)$, while they are larger for a single-tracer analysis if $\ell_{\rm max}=200$. In all cases, the estimated biases are larger for non-overlapping redshift bins. Regarding the use of the Limber approximation, the biases are larger for all parameters  for the multi-tracer case.  The dependence of the estimated biases on the criterion used for $\ell_{\rm max}$ is not significant, except for $\omega_{\rm b}$ and $f_{\rm NL}$. Finally, the estimated biases are again always larger for the non-overlapping redshift bins, with a special mention to  $f_{\rm NL}$, the difference of which is more than $4\sigma$. 

Although exploiting higher multipoles of the angular galaxy power spectra returns smaller biases for the specific sources of theoretical systematic errors explored in this work, we emphasize that there are other key features in the modeling of the observable susceptible to induce a bias in the best-fit parameters. The modeling of non-linear clustering is probably the most important one, and it arises at small scales. Therefore, we advocate for a comprehensive estimation of potential biases in the inferred best-fit parameters using the methodology described in this work, accounting for all possible sources of systematics or approximations adopted, before freezing the analysis pipeline.

\section{Conclusions}
\label{sec:conclusions}
Cosmology needs to transition from the precision to the accuracy era.  Reducing the systematic error budget below the statistical uncertainties represents a crucial step in that direction. Besides controlling observational systematics, improved theoretical models of cosmological observables will be needed. Approximations that have been accurate enough so far, may introduce significant systematic errors for forthcoming experiments.
 
There are two kind of errors that can be introduced into an analysis: a modification of the shape of the posterior distribution and a shift of the location of its peak. These produce a  misestimation of the model parameters covariance and a systematic bias on their best-fit values. While the former has been studied on a companion paper~\cite{Bellomo_Fisher}, here we have focused on the latter.

Expanding upon previous works, we have completely generalized the methodology to estimate the systematic bias introduced in parameter inference when the theoretical model for a given measurement is not accurate enough or the assumed underlying model is not incorrect. The generalized methodology is equally applicable to any measurement, even beyond the field of cosmology.  
 Given the complete flexibility and easy to use of Equations~\eqref{eq:Deltatheta} and~\eqref{eq:systematic} , we advocate  their implementation whenever different approximations are under consideration.

This methodology can also be useful to optimize analyses that rely on an assumed fiducial cosmology. Since this  may introduce a systematic bias (see e.g.,~\cite{Bernal_BAObias}  for a detailed study in the case of the BAO analysis), our methodology can be iteratively applied to find a fiducial cosmology with a prediction more concordant  with the measurements. 
The methodology used in this work assumes Gaussian posteriors since it is based on the Fisher matrix formalism, but this assumption  can be relaxed, following~\cite{joachimi:fishermatriximprovement, Sellentin_dali, sellentin:fishermatriximprovement, amendola:fishermatriximprovement}. This assumption is also dropped when expanding the observable up to higher order on the model parameters. We derive the estimation of the bias when doing a second order expansion in Appendix ~\ref{app:shifts_beyond}.

To show the performance of the generalized methodology, we  have applied it  to the angular galaxy power spectra as observed by next generation galaxy surveys. Instead of considering specific examples, we have used straw-man, yet realistic, galaxy-survey specifications and have shown how neglecting lensing magnification or using the Limber approximation can bias cosmological parameter inference.  We have found significant biases (most of them $\gtrsim 1\sigma$) in all the cases explored in this work. Moreover, we  have also included examples of multi-tracer analyses, using \texttt{Multi\_CLASS}, a modified version of \texttt{CLASS} which allows to compute the angular  cross-power spectra for two different tracers of the matter distribution. In general, we have found that the estimated systematic biases due to the considered approximations are more significant for multi-tracer analyses.

We stress that our results cannot be taken quantitatively at face value because the significance of the biases are expected to be very dependent on the survey and galaxy population specifications. Nonetheless, the risk of introducing significant systematic biases in the results is general, and our work must be considered as a warning for analyses of future experiments. Although, for illustration, we have focused in just two sources of theoretical systematic errors, there are many other potential origins of systematic errors, both observational and theoretical. Therefore, it is of crucial importance that any source of systematics  is reduced well below the $1\sigma$ statistical uncertainty to ensure the robustness of the results. This is because the joint contribution of small systematic errors can significantly bias the inferred constraints otherwise, even if the sources of systematics are unrelated between them.

Considering the dramatic experimental upgrades that many fields of physics will experience in the coming years, it is of paramount importance to fully exploit the potential of new experiments and to obtain unbiased results. To do so, we need to estimate potential biases introduced by approximations under consideration. In this work, we have provided a methodology to do that and estimate the significance of systematic biases introduced in  parameter inference. Finally, we envision that our generalized methodology will also be useful to single out possible sources of systematic errors affecting new or unexpected findings. 

\begin{acknowledgments}
We would like to thank Alan Heavens for comments on last stages of this manuscript. JLB is supported by the Allan C. and Dorothy H. Davis Fellowship, and has been supported by the Spanish MINECO under grant BES-2015-071307, co-funded by the ESF during part of the development of this work. Funding for this work was partially provided by the Spanish MINECO under projects AYA2014-58747-P AEI/FEDER, UE, and MDM-2014-0369 of ICCUB (Unidad de Excelencia Mar\'ia de Maeztu).  NB is supported by the Spanish MINECO under grant BES-2015-073372. AR has received funding from the People Programme (Marie Curie Actions) of the European Union H2020 Programme under REA grant agreement number 706896 (COSMOFLAGS). LV acknowledges support by European Union's Horizon 2020 research and innovation programme ERC (BePreSySe, grant agreement 725327).
\end{acknowledgments}


\appendix
\section{Evaluation of the estimation of the systematic shift and higher order approximations}
\label{app:shifts_beyond}
In Section~\ref{sec:shifts}, we have  presented a general  expression to estimate systematic shifts on the best-fit parameters due to incorrect modeling. Equation~\eqref{eq:Deltatheta} is  obtained  assuming a Gaussian likelihood and expanding the  response of the observables  to small variations in the model parameters up to linear order. In this appendix we evaluate the performance of this approximation, and extend the discussion of Section~\ref{sec:shifts} considering  the expansion up to second order and  for a common non-Gaussian likelihood  corresponding to variables following a Wishart distribution.

\subsection{Accuracy of the estimation}
According to the Fisher forecast approach, the results  for the systematic shift on the best-fit parameters  shown in the main  text   rely on  a linear expansion of the observable on the model parameters (see Equation~\eqref{eq:finidiff}). However, this expansion is not accurate for large displacements $\Delta\mathbf{\theta}$ in parameter space, unless the observable is linear on the model parameters. Moreover, the approach intrinsically assumes a Gaussian posterior.

Here we compare the  shift on the best-fit parameters estimated in that way to those found  from a numerical evaluation of the Gaussian likelihood for the angular galaxy power spectrum. For the sake of simplicity and clarity, we consider a simplified case, in which the only free parameters are $n_{\rm s}$ and $f_{\rm NL}$, and all other parameters are kept fixed. We focus on the bias introduced  by ignoring the contribution of lensing magnification  for an Euclid-like survey with overlapping redshift bins, $\ell_{\rm max}=200$, and  representative fiducial values for the magnification bias parameter $s=\lbrace 0.4,0.6,0.8\rbrace$.  We generate mock data (i.e. angular power spectra)  assuming the fiducial cosmology and including lensing magnification. We do not include sample variance in the mock data, to ease the comparison.
We compute the likelihood of the angular galaxy power spectra for values of $f_{\rm NL}$ and $n_{\rm s}$ in a two-dimensional grid, without including lensing magnification (i.e., assuming $s=0.4$). 

\begin{figure}
\centering
\includegraphics[width=0.8\linewidth]{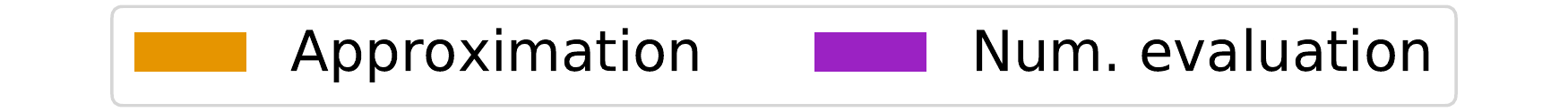}\\
\includegraphics[width=\linewidth]{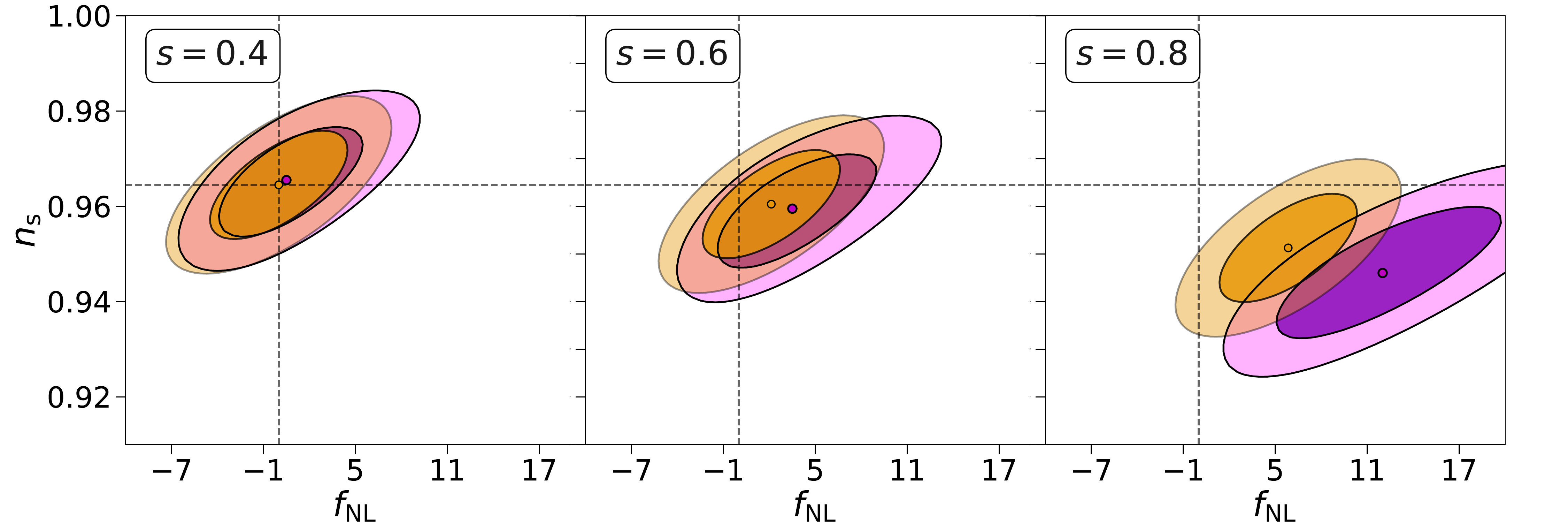}
\caption{Estimated  and numerically evaluated shifts induced by  neglecting lensing magnification with $s=0.4$, $0.6$ and $0.8$ (left, middle and right panel, respectively). Comparison of the forecasted constraints and estimated bias in the best-fit values using the formalism explained in Section~\ref{sec:shifts} (orange) to the results obtained by a numerical evaluation of the likelihood (pink).  We show 68\% and 95\% confidence level constraints for an Euclid-like survey and using $\ell_{\rm max}=200$ for all redshift bins. Dashed grey lines mark the values used to compute the mock data. In the left panel the small discrepancy in the best-fit values  is due to the discrete grid adopted in the numeric evaluation. }
\label{fig:compare}
\end{figure}

The comparison is shown in Figure~\ref{fig:compare}. This figure demonstrates that the estimate of  Equation~\eqref{eq:Deltatheta} (Equation~\eqref{eq:Delta_theta_Cl} for this specific case) and the formalism discussed in Section~\ref{sec:shifts} are qualitatively correct. However, as expected, the accuracy of the estimate decreases as the modelling  error (and hence the bias introduced in the best-fit parameters) increases. In this case, our approach with a linear-order expansion underestimates the bias on the model parameters. Nonetheless, this is very likely to be very case-dependent. For the relevant cases where the bias in the parameter inference is small, the linear approximation provides a good estimates. For other cases, it is possible to use a second order approximation, as explained below. 

It is however important to note that, in general,  if the simple estimate ~\eqref{eq:Deltatheta} predicts a negligile shift, it is safe to adopt the approximation considered. On the other hand, if the simple estimate ~\eqref{eq:Deltatheta} predicts a large shift, even if the shift amplitude is not estimated precisely, it is still a clear indication that the
approximation considered should not be adopted.

\subsection{Systematic shift using a second order expansion}
Let us consider a second order expansion of the  response of the observable $\Psi$  to a variations of the model parameters around  their fidicual values $\bm{\theta}^{\rm fid}$:
\begin{equation}
\hat{\Psi}_i(\bm{\theta}^{\rm fid}+\Delta\bm{\theta}) \approx \hat{\Psi}_i^{\rm fid} + \left(\bm{\nabla}_\theta\hat{\Psi}_i^{\rm fid}\right)^T\Delta\bm{\theta}+\frac{1}{2}\Delta\bm{\theta}^T\left(\mathcal{H}_\theta\hat{\Psi}_i^{\rm fid}\right)\Delta\bm{\theta},
\label{eq:finidiff_2ndorder}
\end{equation}
where  $\mathcal{H}_\theta$ is the Hessian operator (with derivatives with respect to the model parameters), applied to the observable $\hat{\Psi}_i$ and evaluated at $\bm{\theta}^{\rm fid}$. This expansion adds the last term to the linear expansion shown in Equation~\eqref{eq:finidiff}. Now we can proceed as explained in Section~\ref{sec:shifts}: if we substitute this expansion in the logarithm of the  likelihood (Equation~\eqref{eq:logLkl}) and maximize it, we find
\begin{equation}
\begin{split}
\sum_{i,j}& \left[\bm{\nabla}_\theta\hat{\Psi}_i^{\rm fid}+\left(\mathcal{H}_\theta\hat{\Psi}_i^{\rm fid}\right)\Delta\bm{\theta}\right]\left({\rm Cov}^{-1} \right)_{ij}\\
&\left[\Psi^{\rm d}_j -\hat{\Psi}^{\rm fid}_j-\left(\bm{\nabla}_\theta\hat{\Psi}_j^{\rm fid}\right)^T\Delta\bm{\theta}-\frac{1}{2}\Delta\bm{\theta}^T\left(\mathcal{H}_\theta\hat{\Psi}_j^{\rm fid}\right)\Delta\bm{\theta}\right] = 0 \,.
\end{split}
\label{eq:maxLkl_2nd}
\end{equation}
After some algebra we can simplify this expression to 
\begin{equation}
\begin{split}
\sum_{i,j}&\left(\bm{\nabla}_\theta\hat{\Psi}_i^{\rm fid}\right)\left({\rm Cov}^{-1} \right)_{ij}\left(\Delta\hat{\Psi}^{\rm fid}_j\right) = F\Delta\bm{\theta}- \sum_{i,j} \left(\mathcal{H}_\theta\hat{\Psi}_i^{\rm fid}\right)\Delta\bm{\theta} \left({\rm Cov}^{-1} \right)_{ij}  \Delta\hat{\Psi}^{\rm fid}_j + \\
& 
\begin{split}
+ \sum_{i,j}\left[ \left(\mathcal{H}_\theta\hat{\Psi}_i^{\rm fid}\right)\Delta\bm{\theta}\right.& \left({\rm Cov}^{-1} \right)_{ij}\left(\bm{\nabla}_\theta\hat{\Psi}_j^{\rm fid}\right)^T+\\
&\left.+\frac{1}{2}\left(\bm{\nabla}_\theta\hat{\Psi}_i^{\rm fid}\right)\left({\rm Cov}^{-1} \right)_{ij}\Delta\bm{\theta}^T\left(\mathcal{H}_\theta\hat{\Psi}_j^{\rm fid}\right)\right]\Delta\bm{\theta}+
\end{split}\\
&+\sum_{i,j}\left[\frac{1}{2}\left(\mathcal{H}_\theta\hat{\Psi}_i^{\rm fid}\right)\Delta\bm{\theta}\left({\rm Cov}^{-1} \right)_{ij}\Delta\bm{\theta}^T\left(\mathcal{H}_\theta\hat{\Psi}_j^{\rm fid}\right)\Delta\bm{\theta}\right]\, 
\end{split}
\label{eq:maxLkl_2nd_final}
\end{equation}
where each line in the right hand side of the expression corresponds to terms linear, quadratic and cubic on $\Delta\bm{\theta}$, respectively, and we define the difference between the data  and the prediction at $\bm{\theta}^{\rm fid}$ as $\Delta\hat{\Psi}^{\rm fid}\equiv \Psi^{\rm d} -\hat{\Psi}^{\rm fid}$. Equation~\eqref{eq:maxLkl_2nd_final} does not have an analytic solution (which instead exists for a linear expansion, see Equation~\eqref{eq:Deltatheta}), but can be used if a more accurate estimate is needed, as it may be the case for large $\Delta\bm{\theta}$. Note that the presence of $\mathcal{H}_\theta$ involves further computation  of second derivatives. Finally, Equation~\eqref{eq:maxLkl_2nd_final} can be manipulated as described in Section~\ref{sec:shifts} in order to consider the combination of likelihoods.

\subsection{Shift for a non-Gaussian likelihood: Wishart distribution}
\label{app:shift_ng_lkl}
Although the procedure followed in Section~\ref{sec:shifts} to estimate the systematic bias introduced in the best-fit parameters is general for any likelihood, the results  given in Equations~\eqref{eq:Deltatheta} and~\eqref{eq:maxLkl_2nd_final} assume a Gaussian likelihood. While in many applications  the adoption of a Gaussian  likelihood is well justified, this is not always the case. Given that the actual expression for $\Delta\bm{\theta}$ can differ depending on the likelihood, it is important to explore non-Gaussian cases.

As discussed in Paper I, using a Gaussian likelihood for the angular galaxy power spectra is an approximation based on the central limit theorem: the true Gaussian\footnote{Non-linear clustering induces small deviations from Gaussianity, but these are significant only at small (i.e., non-linear) scales.} random variables are the spherical harmonics coefficients associated to the galaxy number  density perturbations. Hence, the angular power spectra follow a Wishart distribution. Since we have used  the angular galaxy power spectra in this work, here we derive $\Delta\bm{\theta}$ for a likelihood of variables following a Wishart distribution.

Let us consider a Gaussian variable $\varphi$ (e.g., the galaxy number density fluctuations), the covariance of which is given by the quantity $\Phi$, considered Gaussian throughout this work, but actually following a Wishart distribution. In most of the real cases analogous to this situation, $\varphi$ is given by the data, and the dependence on the model is encoded in $\hat\Phi$. $\Psi$ (i.e.,  the observable considered in Section~\ref{sec:shifts}) is the half-vectorization of $\Phi$; this is analogous to $\bm{C}_\ell$ and $\mathcal{C}_\ell$, discussed in Section~\ref{sec:observable}.   If we consider a Gaussian likelihood for $\varphi$, neglecting constant terms, we have 
\begin{equation}
\begin{split}
-2\log\mathcal{L}\left(\varphi^{\rm d}\lvert M,\bm{\theta}\right) &  =  \sum_{i,j}\varphi_i^{\rm d}\left(\hat{\Phi}^{-1}(\bm{\theta})\right)_{ij}\left(\varphi_j^{\rm d})\right)^* + \log\left\lvert\hat{\Phi} (\bm{\theta}) \right\rvert =  \\
& = \sum_{p,q}\left[\Phi^{\rm d}_{pq}\left(\hat{\Phi}^{-1}(\bm{\theta})\right)_{qp}\right] + \log\left\lvert\hat{\Phi} (\bm{\theta}) \right\rvert \, ,
\end{split}
\end{equation}
where the  last equality yields the likelihood for $\Phi^{\rm d}$, following a Wishart distribution with the parameter dependence encoded in the matrix $\hat\Phi$. The maximization condition for this likelihood is (dropping the explicit dependence on the parameters and model for the sake of conciseness)
\begin{equation}
\begin{split}
-2\nabla_\theta\log\mathcal{L}  & =\nabla_\theta   \sum_{p,q}\left[\Phi^{\rm d}_{pq}\left(\hat{\Phi}^{-1}\right)_{qp}\right] + \nabla_\theta\log\left\lvert\hat{\Phi} \right\rvert =  \\ & =\sum_{p,q}\left[\Phi^{\rm d}_{pq}\nabla_\theta  \left(\hat{\Phi}^{-1}\right)_{qp} + \left(\hat{\Phi}^{-1}\right)_{pq}\nabla_\theta \hat{\Phi} _{qp} \right] =  \\
& = \sum_{p,q,r,s}\left[-\Phi^{\rm d}_{pq}\left(\hat{\Phi}^{-1}\right)_{qr}\nabla_\theta \hat{\Phi}_{rs} \left(\hat{\Phi}^{-1}\right)_{sp} + \left(\hat{\Phi}^{-1}\right)_{pq}\hat{\Phi}_{qr}\left(\hat{\Phi}^{-1}\right)_{rs}\nabla_\theta \hat{\Phi} _{sp} \right] =   \\
&  = \sum_{p,q,r,s}\left[\left(\hat{\Phi}^{-1}\right)_{pq}\left(\hat{\Phi}_{qr}-\Phi^{\rm d}_{qr}\right)\left(\hat{\Phi}^{-1}\right)_{rs}\nabla_\theta \hat{\Phi} _{sp} \right] = 0 \, ,
\end{split}
\end{equation}
where for any square matrix~$A$,  $\sum_{i,j}A_{ij}A_{ji}={\rm Tr}\left[AA\right]$ (with `Tr' denoting the trace operator), and we have used the following properties:
\begin{equation}
\begin{split}
& \nabla_\theta\lvert A\lvert = \sum_{i,j}\left[{\rm adj}(A)_{ij}\nabla_\theta A_{ji}\right] = \lvert A\lvert \sum_{i,j}\left[\left(A^{-1}\right)_{ij}\nabla_\theta A_{ji}\right] \, , \\
& \nabla_\theta A^{-1} = -A^{-1}\nabla_\theta A A^{-1}\, ,
\end{split}
\end{equation}
where~$\mathrm{adj}(A)$ is the adjugate of the matrix~$A$. Using the same matrix properties as in Equation~\eqref{eq:fisher_likelihood_delta}, we can write the maximization condition as
\begin{equation}
-2\nabla_\theta\log\mathcal{L} \propto \sum_{i,j}\nabla_\theta\hat{\Psi}_i\left({\rm Cov}^{-1}\right)_{ij}\left(\Psi_j^{\rm d}-\hat{\Psi}_j\right)=0\, ,
\end{equation}
which is the same condition as for the Gaussian likelihood of Equation~\eqref{eq:logLkl}. Therefore, at any order in the expansion of the observable on the parameters, the results found in Equation~\eqref{eq:Deltatheta} also apply for the likelihood of Wishart distributed variables. This is because the two likelihoods share the same best-fit parameters, even if the distributions are different. 

\section{Systematic bias as function of the largest scales included}
\label{app:ellmin}
\begin{figure}
\centering
\includegraphics[width=0.8\linewidth]{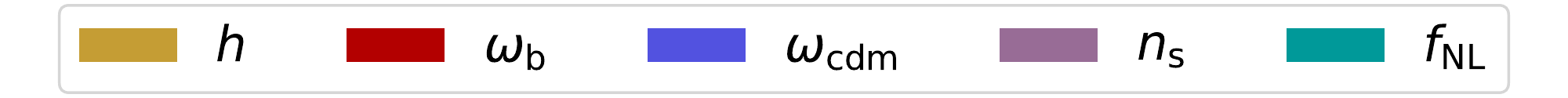}\\
\includegraphics[width=\linewidth]{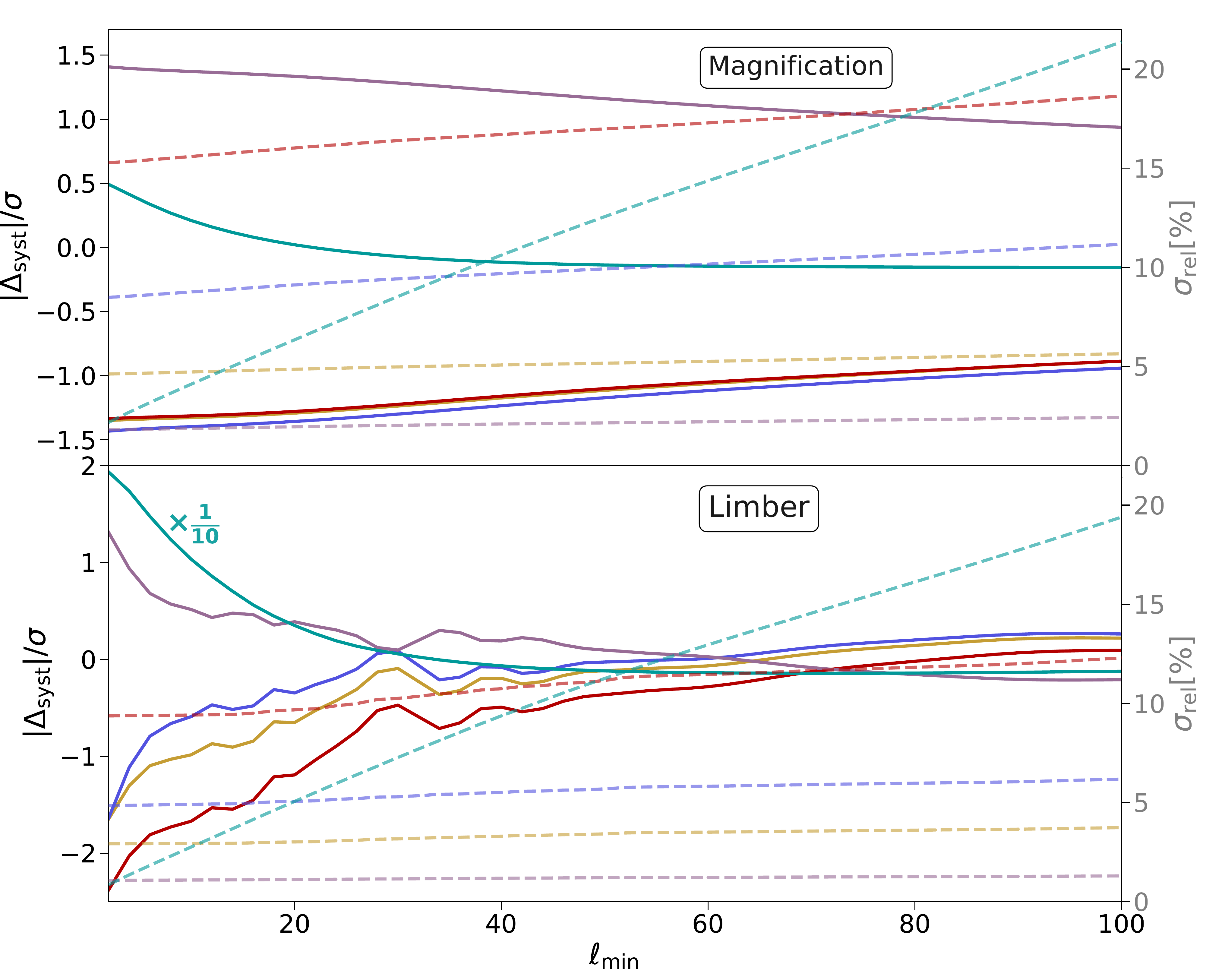}
\caption{Ratio of the estimated bias in the cosmological parameters over the forecasted 68\% confidence level marginalized constraints (solid lines, left side $y$-axis)  compared to the relative forecasted 68\% confidence level marginalized constraints -- except for $f_{\rm NL}$, for which we show the absolute forecasted constraint -- (dashed lines, right side $y$-axis), both of them as function of the minimum multipole $\ell_{\rm min}$ included in the analysis. We show results for the case when the lensing magnification is not modeled (upper panel) and the Limber approximation is used (bottom panel), considering the combination of an Euclid-like survey and a SPHEREx-like survey performing a multi-tracer analysis. In all cases we consider $s=0.6$ for both galaxy populations, and $\ell_{\rm max}(z)$. Note that in the lower panel, the result corresponding to $\Delta_{{\rm syst},f_{\rm NL}}/\sigma_{f_{\rm NL}}$ is divided by a factor 10 for visualization purposes.}
\label{fig:lmin}
\end{figure}

In our analysis, we have found that, in most cases, the bias due to neglecting lensing magnification or using the Limber approximation is more significant when $\ell_{\rm max}=200$ is set (see e.g., Figures~\ref{fig:relshifts_magnification} and~\ref{fig:multi_single}). This is because these two approximations are  increasingly less accurate at low $\ell$. Although we advocate the use of accurate approximations throughout and exploit the whole set of observations (instead of limiting the extent of the analysis to a regime where approximations hold), it is interesting to explore how the significance of the bias decreases as we reduce the multipole range  included in the analysis by  increasing the minimum multipole $\ell_{\rm min}$. The expected reduction of the significance of the bias is due to a smaller $\Delta_{{\rm syst},a}$ -- because the data for which the approximation is less accurate are not included -- and due to an increase in $\sigma_a$ -- given that we are using less data --.

In Figure~\ref{fig:lmin}, we  compare $\Delta_{{\rm syst},a}/\sigma_a$ and $\sigma_a/a$ ($y$-axis legend on the left and right side of each panel, respectively) as a  function  $\ell_{\rm min}$ for our default choice of $\ell(z)$. We show estimations for the five cosmological parameters considered, for the multi-tracer case, and both for the case where lensing magnification is neglected (upper panel) and the Limber approximation is used (bottom panel). For the case in which lensing magnification is neglected, increasing $\ell_{\rm min}$ does not reduce significantly the biases in the inferred parameters, while for the case in which the Limber approximation is used, the systematic errors are below $\sim 0.5\sigma$ for $\ell_{\min}\gtrsim 60$. However, in both cases, the constraining power on $f_{\rm NL}$ (one of the main scientific targets of  the next-generation galaxy surveys) quickly degrades as we increase $\ell_{\rm min}$. 

We want to emphasize that this discussion is mostly for illustration purposes, and that it is certainly desirable to adopt  accurate approximations for the data available, rather than discarding data to be able to use  inferior approximations. Moreover, in the case of pursuing a constraint on $f_{\rm NL}$, using as much data as possible on the largest scales is key.


\bibliographystyle{utcaps}
\bibliography{biblio}


\end{document}